\documentclass{aastex}
\usepackage{emulateapj5}

\newcommand{\n}{\nodata}

\def\gtrsim{\mathrel{\hbox{\rlap{\hbox{\lower4pt\hbox{$\sim$}}}\hbox{$>$}}}}
\def\lesssim{\mathrel{\hbox{\rlap{\hbox{\lower4pt\hbox{$\sim$}}}\hbox{$<$}}}}

\shortauthors{Lister et al.}
\shorttitle{Pearson-Readhead Survey From Space}

\begin{document}
\slugcomment{Accepted for publication in the Astrophysical Journal}
\title{The Pearson-Readhead Survey of Compact Extragalactic \\
 Radio Sources From Space. II. Analysis of Source Properties}
\author{M. L. Lister\altaffilmark{1,2}, S. J. Tingay\altaffilmark{1,3},
 R. A. Preston\altaffilmark{1}}

\altaffiltext{1}{Jet Propulsion Laboratory, California Institute of Technology,
 MS 238-332, 4800 Oak Grove Drive, Pasadena, CA 91109-8099}
\altaffiltext{2}{Current address: NRAO, 520 Edgemont Road,
Charlottesville, VA 22903-2454}
\altaffiltext{3}{Australia Telescope National Facility, P.O. Box 76,
Epping, NSW 2121, Australia}
\email{mlister@nrao.edu}

\begin{abstract}
We have performed a multi-dimensional correlation analysis on the
observed properties of a statistically complete core-selected sample
of compact radio-loud active galactic nuclei, based on data from the
VLBI Space Observing Programme (Paper I) and previously published
studies. Our sample is drawn from the well-studied Pearson-Readhead
(PR) survey, and is ideally suited for investigating the general
effects of relativistic beaming in compact radio sources. In addition
to confirming many previously known correlations, we have discovered
several new trends that lend additional support to the beaming
model. These trends suggest that the most highly beamed sources in
core-selected samples tend to have a) high optical polarizations; b)
large pc/kpc-scale jet misalignments; c) prominent VLBI core
components; d) one-sided, core, or halo radio morphology on kiloparsec
scales; e) narrow emission line equivalent widths; and f) a strong
tendency for intraday variability at radio wavelengths.  We have used
higher resolution space and ground-based VLBI maps to confirm the
bi-modality of the jet misalignment distribution for the PR survey,
and find that the sources with aligned parsec- and kiloparsec-scale
jets generally have arcsecond-scale radio emission on both sides of
the core.  The aligned sources also have broader emission line
widths. We find evidence that the BL Lacertae objects in the PR survey
are all highly beamed, and have very similar properties to the
high-optically polarized quasars, with the exception of smaller
redshifts. A cluster analysis on our data shows that after partialing
out the effects of redshift, the luminosities of our sample objects in
various wave bands are generally well-correlated with each other, but
not with other source properties.
\end{abstract}


\section{Introduction}

One of the major obstacles to a better understanding of active
galactic nuclei (AGNs) is the disentanglement of orientation and
relativistic beaming effects on their observed properties. The use of
high-resolution very long baseline inteferometric (VLBI) techniques
has led to enormous progress in this regard, but the determination of
viewing angles and intrinsic speeds of relativistic jets in compact
extragalactic radio sources remains exceedingly
difficult. Considerable effort has gone into the study of individual
objects (e.g., 3C 279: \citealt{WPU00}; 3C 345:
\citealt{LZ99}), but it is unclear how representative these sources
are of the overall radio-loud AGN population. As a result, the degree
to which relativistic beaming biases our view of compact radio sources
is not well known.

An effective method of addressing these issues is through studies
of large well-selected samples of objects, of which several are
currently underway \citep[e.g.,][]{TVRP96, Kell98, FHM00}. In a companion paper
(\citealt{L-PRI}; Paper I), we describe one such sample drawn from
the Pearson-Readhead (PR) survey for which we have obtained
high-quality 5 GHz space-VLBI images with the VLBI Space Observing
Programme (VSOP).  The PR survey is arguably one of the best-studied
radio samples in existence, with an enormous amount of supporting data
available at a variety of wavelengths. As such, it is ideally
suited for studying the effects of relativistic beaming on a
wide variety of source properties, and for testing general predictions
of the beaming model. 

Here we carry out a multi-dimensional correlation analysis of the
parsec- and kiloparsec-scale properties of a flat-spectrum,
core-selected subset of the Pearson-Readhead survey.  We identify
several new trends involving radio morphology and intraday variability
which suggest that these  properties are highly influenced by
relativistic beaming. We also confirm previous results which show that
the core dominance, optical polarization, and emission line equivalent
width are useful statistical beaming indicators.  In \S2 we describe
our sample selection criteria, followed by a description of the
observed source properties in \S3. We describe our statistical tests
in \S4. In \S\ref{cluster} we use cluster analysis to investigate
natural groupings of source properties, and discuss the role of
emission line strength as a beaming indicator in
\S\ref{BL_LAC}. In Sections \ref{simulations} and \ref{recognition} we
describe Monte Carlo beaming simulations and possible ways to identify
highly beamed sources. Finally in \S6 we summarize our results and
make suggestions for future work.

\section{\label{sample_desc} Sample description}
In Paper I we described a subsample of 31 compact objects from the
Pearson-Readhead survey that were chosen as suitable targets for
space-VLBI. These objects all have flux densities $>0.4$ Jy on the
longest Earth baselines ($\sim 10,000$ km) at 5 GHz. This sample
(hereafter referred to as VSOP-PR) is well-suited for ensuring likely
fringe-detection on space baselines, but is not statistically complete
from the standpoint of relativistic beaming. Since an accurate
relation between intrinsic core and lobe emission has not yet been
established, current beaming simulations can only make predictions for
the properties of AGN core components. Therefore, one needs a sample
that is selected on the basis of core flux only. The simplest way to
obtain such a sample is to use a high selection frequency
(e.g. $\gtrsim 15$ GHz), at which the steep-spectrum lobe emission
makes a negligible contribution to the total flux density. However,
currently available all-sky radio surveys only range up to 5 GHz. As a
result, nearly all large radio surveys (such as the Pearson-Readhead
sample) tend to contain a mixture of core- and lobe-dominated objects.

A common method of excluding the lobe-dominated objects from radio
surveys is to adopt a simple two-point spectral-index criterion, such
that all objects with spectra steeper than a certain value (usually
$\alpha < -0.5\; ; \  S\propto \nu^{\alpha}$) are excluded. This method has
some drawbacks, as the spectral indices may not always be based on
simultaneous flux density measurements, and can be affected by source
variability. Also, some sources whose spectra peak around a few GHz (the
so-called Gigahertz Peaked Spectrum, or GPS sources) may have flat
two-point spectral indices, and will be included in the final
sample. Recent evidence has indicated that these GPS sources, along
with a similar class, the compact steep-spectrum sources, are
much younger and more luminous than typical compact AGNs,
and have weak parsec-scale core components that are not likely to be
highly beamed \citep{TRP96}. On this basis they should be excluded
from core-selected samples.

In consideration of these issues, we have constructed a core-selected
subset of the Pearson-Readhead survey by using time-averaged 5-15 GHz
spectral indexes from the University of Michigan monitoring program,
as tabulated by \cite{AAH92}. These were computed from monthly
averages of measurements taken over several years, and are less
susceptible to the variability errors described above. By adopting a
cutoff of $\bar\alpha_{5-15} > -0.4$ we effectively eliminated all of
the GPS and compact steep spectrum sources from the PR sample. Our
final flat-spectrum Pearson-Readhead sample (hereafter referred to as
the FS-PR) contains 32 objects, 28 of which are members of the VSOP-PR
sample. The four additional objects are the quasars 0723+679,
0850+581, 0954+556, and 2351+456, and are targets of upcoming
space-VLBI observations with VSOP.

\section{Observed Source Properties\label{props}}
We have gathered a vast amount of data on the FS-PR and VSOP-PR
samples from our space VLBI data presented in Paper I and previously
published studies. A summary of these quantities is given in
Table~\ref{quantities}, and a detailed explanation follows in
\S\ref{gendesc}. Throughout this paper we use a standard Freidmann
cosmology with $H_o = 65 \rm \; km \; s^{-1}
\;Mpc^{-1}$, $q_o = 0.1$, and zero cosmological constant.  We give all
position angles in degrees east of north, and define the spectral
index according to $S_\nu \propto \nu^\alpha$.

\subsection{\label{gendesc} General properties}
In Table ~\ref{genprops} we list various general properties of our
sources. In columns 1 and 2 we give the IAU and common name of each
source, and indicate whether it is a member of the third EGRET catalog
of gamma-ray loud sources \citep{HBB99}. In column 3 we give the
optical classification as a radio galaxy (RG), BL Lac object (BL), or
quasar (Q). If a particular quasar has a published optical
polarization greater than $3\%$ at any epoch we classify it as a
high-optically polarized quasar (HPQ). For the remaining quasars, we
classify those with three or more polarization measurements with
$m_{opt} < 3 \%$ as low-optically polarized radio quasars (LPRQ). Our
BL Lac classifications are based on the catalog of \cite{PG95}.

We have used previously published arcsecond-scale radio images to
classify our sources into four basic morphological types depending on
whether they possess distinct hotspots on one or both sides of the
core (column 4). Those sources with no hotspots are termed either
compact or halo, with the latter displaying weak diffuse structure
surrounding the core component.

\subsection{\label{variabdesc} Variability properties}
\cite{AAH92} have monitored the long-term variability properties of
the PR sample, and defined a peak-to-trough variability index V: 

\begin{equation}
V= {{(S_{max}-\sigma_{Smax}) - (S_{min} + \sigma_{Smin})} \over
{(S_{max}-\sigma_{Smax}) + (S_{min} + \sigma_{Smin})}},
\end{equation}

where $S_{max}$ and $S_{min}$ are the highest and lowest flux
measurements, and the $\sigma$'s are the associated measurement
errors. We will refer to their published indices at 4.8 and 14.5 GHz in
our study as $V_5$ and $V_{15}$ respectively.

A large number of Pearson-Readhead sources have also been monitored
for intraday variability (IDV) with the 100m Effelsberg telescope
and/or the VLA\footnote{The Very Large Array (VLA) is maintained by
The National Radio Astronomy Observatory, which is a facility of the
National Science Foundation operated under cooperative agreement by
Associated Universities, Inc.}
\citep{Q00}.  These authors have classified sources into three
distinct IDV classes on the basis of their structure functions. Type
II objects exhibit significant up-and-down variations on hourly
timescales, while type I objects show monotonic rises or decays in
flux on timescales less than a day. Type 0 objects show no apparent
IDV activity. Statistical studies have shown that a particular source
may have different IDV classifications at different times. For those
PR sources that have been monitored on one or more occasions for IDV,
we list in column 5 of Table~\ref{genprops} the highest IDV
classification from \cite{KQW92},
\cite{Q92}, and \cite{Q00}.
 
\subsection{Optical spectral properties}
\cite{LZR96} have obtained high-quality optical spectra of  all
the Pearson-Readhead survey objects with the 5m Palomar telescope. In
column 6 we list their measured redshifts, and in column 7 we list the
rest-frame equivalent width of the widest permitted emission line in \AA,
corrected to the source frame.

\subsection{Luminosity and core properties}
In column 8 of Table~~\ref{genprops} we list the 5 GHz luminosity of
the parsec-scale core component, measured in the source rest frame
assuming $\alpha_{core} = 0$. The core flux density and corresponding
rest frame brightness temperature are given in Paper I.

In column 9 we list the total luminosity of the source at 5 GHz, also
in the source frame. We determined the total flux density and 5-15 GHz
spectral index at the VSOP observation epoch (where applicable) by
interpolating the University of Michigan radio light
curve\footnote{http://www.astro.lsa.umich.edu/obs/radiotel/umrao.html}
for each source. For those sources where 15 GHz data weren't
available, we used the time-averaged spectral index listed in
\cite{AAH92}. 

The relative prominence of the unresolved VLBI core with respect to
extended emission in an AGN is a useful indicator of relativistic
beaming, since the latter emission is thought to be largely isotropic
and not highly beamed. There is no standard way of defining this
parameter in the literature, with different authors adopting various
ratios of core-to-total flux, core-to-extended flux, or total
parsec-scale to total (single-dish) flux. In this paper we adopt a
parsec-scale core dominance ($R_{pc}$), which we define as $L_{core}/
(L_{tot} - L_{core})$, where $L_{tot}$ is the total (single-dish)
luminosity.  We list the logarithm of this quantity (in the source
frame) in column 10 of Table~\ref{genprops}.

A beneficial aspect of the PR sample is that high-quality
arcsecond-scale 1.4 GHz images have been obtained for nearly all of
the sources as part of the Caltech-Jodrell surveys \citep{PWR95} or
other studies.  In columns 11 and 12 of Table~\ref{genprops} we list
the 1.4 GHz luminosities of the arcsecond-scale core and extended
structure, respectively, as measured from VLA or MERLIN images. The
majority of these data are from \cite{MBP93}, who obtained high dynamic range
images of a large sample of core-dominated AGNs with the VLA. Here we
assume typical spectral indices of $\alpha = 0$ for the core
components, and $\alpha = -0.8$ for the extended emission. For those
objects with no detectable extended emission, we assumed an upper
limit of three times the rms noise level in the map. This limit is
valid provided the undetected emission is spread over an area larger
than the restoring beam.

In column 13 we tabulate the kiloparsec-scale core dominance ratios
($R_{kpc}$) for our objects, which is simply the ratio of core to
extended luminosity at 1.4 GHz, measured in the rest frame of the
source. In column 15 we list the monochromatic luminosities of our
sources in the 1 keV X-ray band. These were derived from fluxes in the
literature, assuming $\alpha_{xray} = -0.8$.

We have also calculated optical luminosities (not listed) based on
optical V band magnitudes from
NED\footnote{http://nedwww.ipac.caltech.edu/}. We made no correction
for galactic absorption, and adopted a standard photometric
transformation \citep{W81} and an optical spectral index $\alpha =
-0.5$.

\subsection{Maximum apparent jet velocity}
The apparent velocities of distinct components in AGN jets,
combined with estimates of the viewing angle, can provide useful
information on the bulk Lorentz factor of the flow. Many superluminal
AGNs (e.g., 3C~345; \citealt{RZL00}) exhibit components moving at
different velocities, which suggests that we are likely seeing a pattern
speed rather than the true speed of the underlying
plasma. Nevertheless, the maximum observed velocity can still provide a
useful constraint on the amount of relativistic beaming in a
jet. Apparent velocity data are available for 23 sources in the
VSOP-PR and FS-PR samples. In column~2 of Table~\ref{jetprops} we list
the maximum velocity for these sources in units of $c$.

\subsection{\label{bend_meas} Jet bending}
There is a great deal of interest in mapping the apparent trajectories
of AGN jets at high resolution, as small variations in viewing angle
can dramatically alter the apparent brightness distribution along a
relativistic jet \citep{GAM94}. Many authors (e.g., \citealt{PR88};
\citealt{TME98}) have used $\Delta PA$, the difference between the jet
position angles measured on milliarcsecond and arcsecond scales, as a
simple measure of jet bending. For those sources that have either
straight jets or a single bright component on arcsecond scales, the
measurement of $\Delta PA$ is straightforward.  However, as pointed
out by \cite{BGP98}, the wide variety of complex curvatures and
morphologies found in AGN jets makes it difficult to measure $\Delta
PA$ in a uniform manner for large samples. For some curved jets such
as 1633+382, there have been up to five different values of $\Delta
PA$ published in the literature.  We have attempted to limit possible
biases due to redshift in our analysis by defining the position angle
on kiloparsec scales ($PA_{kpc}$) to be the position angle of the
arcsecond-scale jet at a projected distance of $10\; h^{-1}$ kpc from
the core. In cases where there was no identifiable jet, we used the
position angle of the first radio component located farther than this
projected distance. We believe this is the most self-consistent
method, given the large range of redshift range (and in turn, image
surface brightness) of our sample.

We define the position angle on parsec scales ($PA_{pc}$), as the
position angle of the innermost jet component with respect to the
core.  Since the region near the core in most of our sources is still
optically thick at 5~GHz, we are currently obtaining 43 GHz images of
the FS-PR sample with the VLBA to obtain better estimates of $PA_{pc}$
(Lister et al., in preparation).  In column 2 of Table~\ref{jetprops}
we list the innermost jet position angle at 43~GHz for those sources
which we already have data, otherwise we give a position angle based
on our 5~GHz VSOP image or other published VLBI data. We list the
$\Delta PA$ values for our sources in column 8 of
Table~\ref{jetprops}.

\subsection{Polarization properties}
Extensive data on the linear polarization properties of the PR sample
in the optical and radio regimes have been tabulated by \cite{ILT91}
and \cite{AAH92}, respectively. In Table~\ref{quantities} we list the
measured properties from these papers that we include in our
correlation analysis. Our optical polarization data are taken from
\cite{ILT91} with the exception of 1954+513, whose data are from
\cite{WWB92}. We introduce the derived quantity $|\chi_{15} -
PA_{pc}|$, which represents the angular offset between the inner jet
position angle and the electric polarization vector angle ($\chi$) at
15 GHz. The $\chi_{15}$ values are time-averaged single-dish
quantities tabulated by \cite{AAH92}. Due to the $180\arcdeg$
ambiguity in the $\chi$ measurements, $|\chi_{15} - PA_{pc}|$ cannot
exceed $90\arcdeg$.

\section{Correlation analysis}
\subsection{\label{stat_tests} Description of statistical tests}
The large variety of data that we have gathered on the FS-PR sample
are well suited for a multi-dimensional correlation analysis. This
method checks for correlations among all pairwise combinations of
variables, and is therefore unbiased with respect to any {\it a
priori} expectations.  A major caveat, however, is the possibility of
detecting spurious correlations. For example, when testing for
correlations among $n$ variables, if one chooses a significance level
$ p > P$ to call a significant correlation, you should expect to find
$(1-P)n(n-1)/2$ spurious correlations. Unfortunately, it is usually
impossible to determine which correlations are spurious, and which are
genuine. If the dataset is large enough, it is sometimes feasible to
run correlation tests on different subsets of the data. Alternatively,
the tests can be run on an entirely new dataset. Neither of these
options is available to us for the FS-PR sample, so we have reduced
the chance of false detections by adopting a relatively high
confidence level of $98\%$. For every correlation above this
significance, we also checked the scatter plot to determine whether
the significance level had been artificially increased by the presence
of an outlying data point. If the significance dropped below $(98\%)$
after the removal of the outlier, we rejected the correlation.

To determine the correlation significance among pairs of variables, we
employed a Kendall's tau test. This test does not require that the
intrinsic relation between variables be linear (i.e., it is
non-parametric), and considers only the relative ranks of individual
data pairs. Since many of our data points are censored (i.e., are
limits only), we employed a special version of Kendall's tau that
incorporates the technique of survival analysis. We used the
algorithms provided in the ASURV package \citep{ASURV} for this
purpose.

Many of the measured properties of flux-limited AGN samples have a
large dependence of redshift, due to the steepness of the parent
luminosity function. Intrinsically faint radio sources in these
samples typically have low redshifts due to the flux cutoff. As as
result, certain variables with a mutual dependence on redshift, such
as luminosities at different wavelengths, can appear to be very
well-correlated, even though no intrinsic correlation may exist (see,
e.g., \citealt{Pad92}).  In order to eliminate these redshift-induced
correlations from our analysis, we calculated partial correlation
coefficients of the form

\begin{equation} \label{partcorr}
\tau_{ab,z} = {\tau_{ab} - \tau_{az}\tau_{bz} \over
{\sqrt{1-\tau_{az}^2}\sqrt{1-\tau_{bz}^2}}}, 
\end{equation}

where the $\tau $'s represent the various Kendall's tau correlation
coefficients between variables $a$,$b$, and redshift ($z$). Extra
steps must be taken when calculating partial correlations involving
censored data. We used the algorithms of \cite{AS96} that have been
developed for this purpose.

In addition to our Kendall's tau tests, we also performed
statistical tests to look for differences in source properties among
different classes, e.g., BL Lacs versus quasars and EGRET versus non-EGRET
sources. For those variables that involved limits, we used Gehan's
generalized Wilcoxon two-sample test \citep{G65} from the ASURV
package. For the remaining variables we used a standard
Kolmogorov-Smirnov (K-S) test. In both cases we adopted a $98\%$
confidence level in determining whether the differences were
significant. 

\subsection{\label{stat_results} Results}
Our Kendall's tau tests have identified 30  out of 171 possible
variable pairs that are correlated above the $98\%$ confidence level
(we discuss additional correlations with redshift separately in
\S\ref{redshift}). From our previous discussion in \S4.1, we can
expect approximately three of these to be due to chance. Of the
original 30 correlations, we reject six as likely being due to a
mutual dependence on redshift, and another three as a result of one or
two outlier points that artificially boost the confidence level. This
leaves 21 bone fide correlations for the sample, 20 of which are
either expected or have been previously found in other studies (see
Table~\ref{corr_results}).

We have resorted to several different methods in order to summarize
the complex inter-relations among this large collection of source
properties. Since a single correlation matrix is too large to present
here, we list in Table~\ref{corr_results} only the 30 initial pairs
that are significant at the $98\%$ level or higher.  We give a
reference for each correlation that has been previously detected for
radio-loud AGNs by other authors. We note that in some cases this may
not refer to the very first detection of a particular correlation.

In Table~\ref{ks_results} we summarize the results of our two-sample
tests. The values represent the probabilities (in per cent) that the
samples were drawn from the same parent population. For clarity, we
have omitted all values from this table that exceed $2\%$.

\begin{figure*}
%\epsscale{0.6}
\plotone{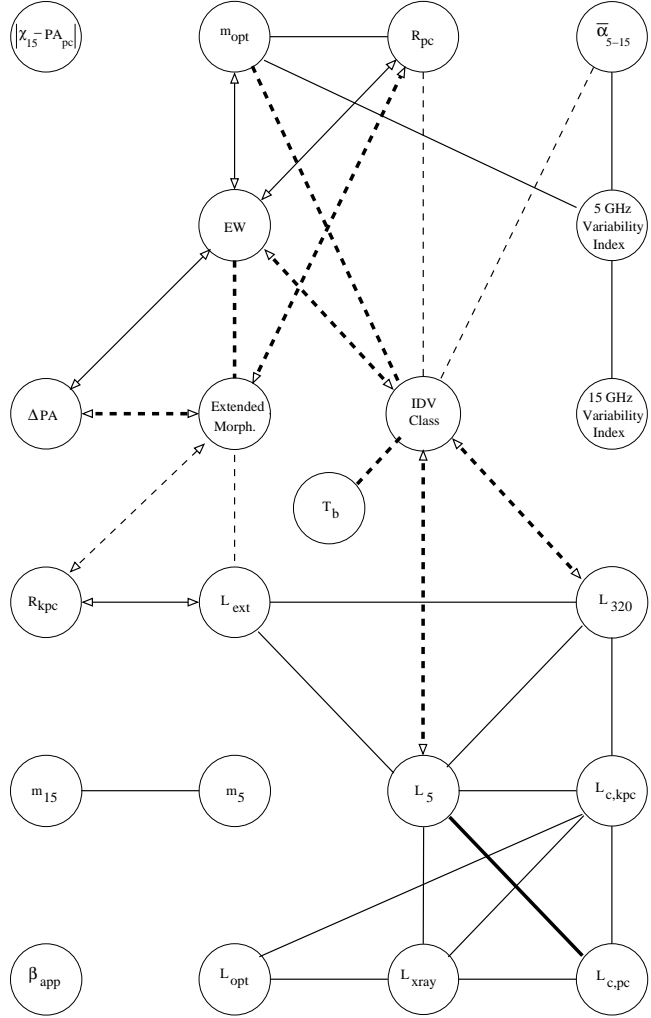}
\caption{\label{network} Diagram summarizing interrelations among source
properties for the FS-PR sample (see Table~\ref{quantities} for symbol
definitions). Solid lines indicate correlations found from Kendall's
tau tests. Dashed lines indicate significant population differences
according to two-sample tests. Arrows denote anti-correlations, and
lines in bold represent new correlations detected in this study. }
\end{figure*}

In Figure~\ref{network} we present a graphical representation of our
test results. The solid lines represent correlations found from
Kendall's tau tests, while the broken lines denote those from the
two-sample tests. Arrows on the lines denote anti-correlations, and the lines
in bold represent new correlations detected in this study. For
example, the dashed line connecting IDV and $m_{opt}$ indicates that
IDV type I and II sources have higher optical polarization levels than
type 0 sources. There are indications of several groupings of
variables in this diagram, which we will discuss in more detail in
\S\ref{cluster}.

% - component luminosities and Tbs
% - say that Tbs are discussed in apj letter
% - section on jet curvature - bend and re-align ?

\subsubsection{\label{redshift} Correlations with redshift}
All of the luminosity variables in our dataset, with the exception of
extended luminosity ($L_{ext}$), are correlated with redshift at
better than $99.999\%$ confidence. This is expected, given the
flux-limited nature of the FS-PR sample.  None of the other
(non-luminosity) variables are significantly correlated with
redshift. We note that the BL Lacertae objects have significantly
lower redshifts than the quasars in the sample
(Table~\ref{ks_results}).

\subsubsection{\label{lum_corr} Correlations involving luminosity variables}

The various luminosities of our sample objects are generally
well-correlated with each other, but not with other source properties,
as is indicated by their location in the lower right-hand corner of
Figure~\ref{network}. One major exception is IDV activity. Our
two-sample tests indicate that sources that have not shown IDV tend to
be more luminous at 320 MHz and 5 GHz. These trends may be due to
beaming and selection effects in the sample (see \S\ref{simulations}).

Among luminosity-luminosity pairs, correlations involving the total 5
GHz luminosity ($L_5$) are the most prevalent. With the exception of
$L_5$ versus the VLBI core luminosity ($L_{c,pc}$), all of the
luminosity-luminosity correlations in Table~\ref{corr_results} have
been previously detected in other studies. The strong correlation
between $L_5$ and $L_{c,pc}$ is merely a reflection of the highly
core-dominated nature of the FS-PR sample.

\subsubsection{Trends with polarization}

The level of polarization at optical wavelengths has historically been
an effective method of identifying highly beamed blazars whose jets
are seen nearly end-on, since very few optically-selected quasars have
$m_{opt} > 3\%$ \citep{SMA84}. In Table~\ref{corr_results} we confirm
the previously known correlations with parsec-scale core dominance, 5
GHz radio variability amplitude, and emission line equivalent width
that have helped establish $m_{opt}$ as a beaming indicator (see,
e.g., \citealt{ILT91}). 

\begin{figure*}
\plotone{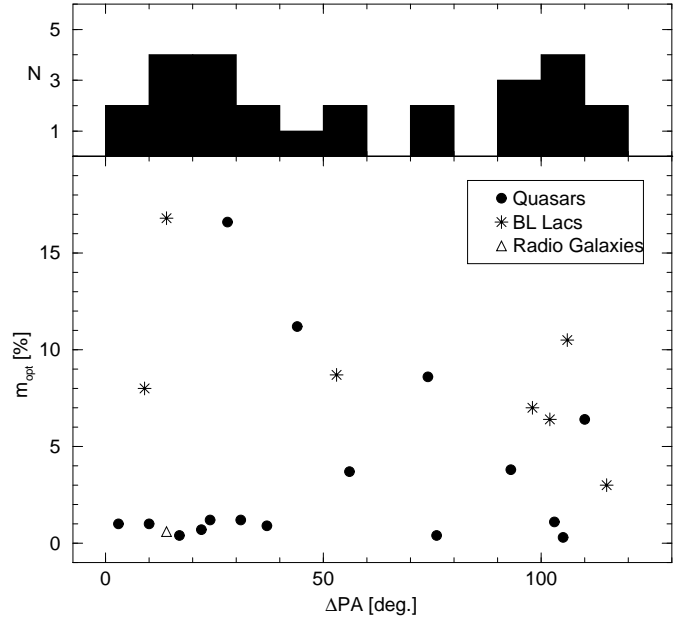}
\caption{\label{mopt__deltaPA} Top panel: histogram of pc/kpc-scale
jet misalignment angle for the FS-PR sample. Bottom panel: jet
misalignment angle plotted against optical polarization for various
optical classes.}
\end{figure*}

\cite{XRP94} found that in the combined PR/Caltech-Jodrell samples, 
sources with low optical polarization levels $(< 3\%)$ tend to have
better jet alignments between parsec and kiloparsec scales than highly
polarized sources. We find some indication of differences of these
groups in the $m_{opt}$ - $\Delta PA$ plane
(Fig.~\ref{mopt__deltaPA}). There is a distinct grouping of sources in
the lower left-hand portion of the plane which have both low optical
polarization and small jet misalignments. The remainder of the sample
shows a roughly linear trend of decreasing optical polarization with
increasing jet misalignment.

It is unclear whether the differences between the high and
low-optically polarized objects in Fig.~\ref{mopt__deltaPA} can be
entirely attributed to beaming and projection effects, which would
tend to amplify any small intrinsic jet bends in sources seen nearly
end-on.  The current consensus is that blazars appear highly polarized
since their optical synchrotron radiation is significantly boosted
with respect to the underlying (thermal) continuum (the same process
is responsible for lowering the observed emission line widths of
highly beamed objects, e.g., \citealt{WWB92}).  If the optical
synchrotron component is not highly polarized to begin with, however,
the optical polarization will remain low, regardless of how beamed the
source is. For example, \cite{LS00} found evidence that many of the
differences between HPQs and LPRQs are not due to differences in
Doppler factor or jet orientation, but rather intrinsic differences in
their jet magnetic field structures. It is possible that blazars go
through temporary quiescent phases of low intrinsic polarization and
flux variability, in which they appear as LPRQs \citep[e.g.,][]{F88}.
Long-term optical polarization monitoring studies of samples such as
the FS-PR are needed to establish the properties of this duty cycle in
blazars.

\subsubsection{\label{IDV} Properties of IDV sources}

In an accompanying paper \citep{TLP00}, we discuss how the PR sources
that have displayed type I and II IDV have substantially higher VSOP
core brightness temperatures than those sources that have not
displayed any IDV. Although it appears that the presence of a bright,
compact core on VLBI scales is a necessary criterion for IDV (e.g.,
\citealt{HKS87,Q92}), the phenomenon only occurs in $\lesssim 2/3$ of
all flat-spectrum radio sources \citep{Q92,KWK99}. Based on the IDV
statistics of a large sample of AGNs, \cite{Q92} concluded that IDV
was more prevalent in sources with $80\%$ or more of their total 5 GHz
flux density contained in an unresolved VLBI core. We detect a similar
trend in the FS-PR, with the $R_{pc}$ distribution of the type 0
(non-IDV) sources differing from that of the type II sources at the
$99.918\%$ confidence level. The type 0 sources also have steeper
radio spectral indices, in agreement with the findings of \cite{Q92}.

We have also detected two additional trends that confirm the
association of IDV with the blazar class. The IDV type I and II
sources have higher optical polarization levels and narrower emission
line equivalent widths than the type 0 objects. This suggests that the
weak-lined BL Lac objects may be the best candidates for IDV;
indeed, an optical study by \cite{HW96} found IDV to be present in 28
of 34 BL Lacs selected from the 1 Jy catalog
\citep{KWP81}. 

\begin{figure*}
\plotone{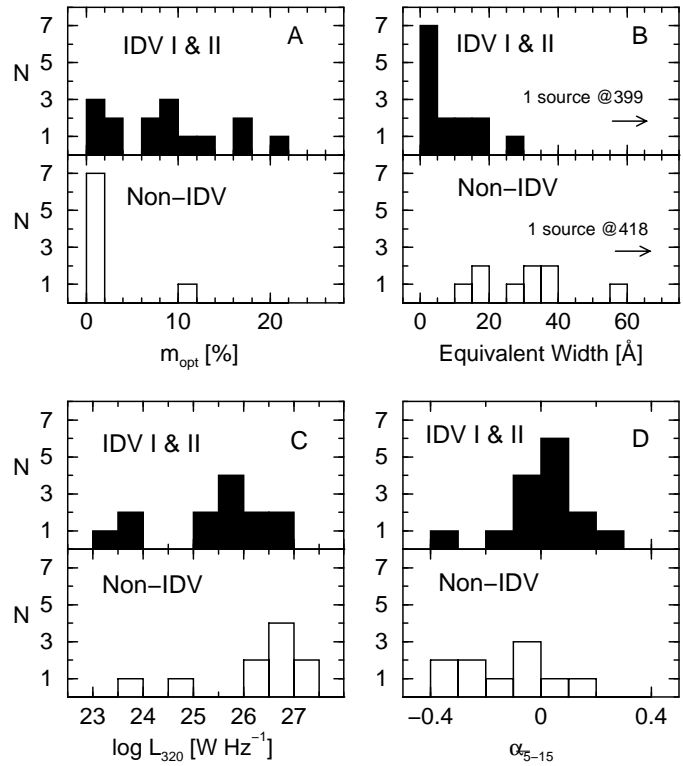}
\caption{\label{IDVhist} Histograms of various properties for IDV type I \& II
(shaded) and non-IDV sources (un-shaded). Panel (A): optical
polarization; (B): source frame equivalent width of widest permitted
line; (C): luminosity at 320 MHz. (D): time-averaged spectral index
between 5 and 14.5 GHz.}
\end{figure*}

The correlations with optical polarization and equivalent width
suggest that the occurrence of IDV depends more on the Doppler factor
of the source rather than any intrinsic property such as 
core size.  As we will show in
\S\ref{simulations}, the highest Doppler factor sources in a
flux-limited sample such as the FS-PR are predicted to have low
intrinsic (unbeamed) luminosities, according to the relativistic
beaming model. Since the 320 MHz luminosity is generally accepted to
be a good indicator of total jet power and intrinsic luminosity
(\citealt{RS91, KA97}), this would imply that the type I and II IDV
sources in the FS-PR should have lower $L_{320}$ values, which is in
fact the case (Fig.~\ref{IDVhist}).

It is significant that we found no trends between IDV and beaming
indicators on kiloparsec scales (i.e., $\Delta PA$, $R_{kpc}$, and
morphological classification).  This further implies that it is the
Doppler factor of the {\it innermost region} of the jet that determines
whether a source will display IDV.

\subsubsection{Dependence of source properties on extended radio
structure\label{morph}} 
\begin{figure*}
\plotone{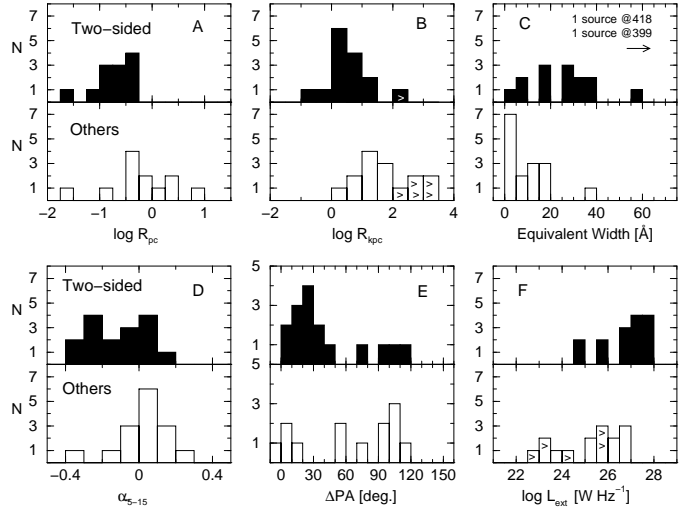}
\caption{\label{morph_hist} Histograms of various properties for sources with two-sided
arcsecond-scale radio morphology (shaded) versus compact, one-sided
and halo sources (un-shaded). Panel (A): parsec-scale core dominance
ratio; (B): kiloparsec-scale core dominance ratio; (C): source frame
equivalent width of widest permitted line; (D): time-averaged spectral
index between 5 and 14.5 GHz; (E): pc/kpc-scale jet misalignment
angle; (F): arcsecond-scale extended luminosity at 1.4 GHz.}
\end{figure*}

Our correlation analysis has revealed that the arcsecond-scale radio
morphology of core-dominated radio sources is closely connected with 
several other important source properties.  In
\S\ref{gendesc} we classified our sources as either compact,
one-sided, two-sided, or halo based on their extended arcsecond-scale
radio structure.  \cite{Mur88} analyzed the kiloparsec-scale
properties of a large sample of core-dominated AGNs, and found that
the two-sided sources tend to have lower core-dominance ratios
($R_{kpc}$) and stronger extended luminosities. We confirm these
findings for the FS-PR, and additionally find that the two-sided
sources have smaller {\it parsec}-scale core dominance, larger
emission line equivalent widths, and smaller jet misalignments between
parsec- and kiloparsec-scales (Fig.~\ref{morph_hist}). \cite{ILT91}
found that the latter three properties are all highly influenced by
source orientation, which implies that the jet axes of the two-sided
sources are oriented farther away from the line of sight than the other
sources in the FS-PR. It is conceivable that the two-sided sources
might have intrinsically smaller core dominance and stronger extended
power (irrespective of beaming), but we consider it unlikely that
intrinsic differences could also conspire to mimic the trends with
equivalent width and jet misalignment.  Furthermore, there is strong
evidence from unbeamed samples such as the 3C survey that the vast
majority of powerful radio loud AGNs are intrinsically two-sided
\citep{HUO83}. A simple explanation for the trends we see with
sidedness is that for sources with jets seen nearly end-on, the
emission from the hotspot on the far side of the one-sided sources is
either beamed away from us, or is swamped by the highly beamed core in
images of limited dynamic range. 

\subsubsection{Jet misalignment and bending}

The difference in inner jet position angle between parsec and
kiloparsec scales ($\Delta PA$) has been used in many studies as a
general indicator of jet orientation. Since AGN jets are known to be
intrinsically bent, those jets that are closer to the line of sight
should have more exaggerated bends due to projection effects, and
usually (but not necessarily) larger values of $\Delta PA$. \cite{PR88}
found unexpectedly that the distribution of $\Delta PA$ for the PR
survey was bi-modal, with a clear division at approximately 45
degrees. \cite{XRP94} found a similar result for the combined
PR/Caltech-Jodrell samples. The reasons for this bi-modality are still
unclear. \cite{CM93} have suggested that it could be due to selection
effects associated with relativistic beaming and jet trajectories in
the form of damped helices.

In the top panel of Figure~\ref{mopt__deltaPA} we show the
distribution of $\Delta PA$ for the FS-PR sample, based on our
improved parsec- and kiloparsec-scale jet position angle measurements
described in \S\ref{bend_meas}. The bi-modality noticed by 
\cite{PR88} is still present, with a division between aligned and
misaligned sources occurring at roughly $60\arcdeg$. The vast majority
of the aligned population ($\Delta PA < 60\arcdeg$) have two-sided
radio structure on arcsecond-scales (Fig.~\ref{morph_hist}).  This
dependence of jet misalignment on lobe morphology strongly suggests
that the highly misaligned sources have inner jets that are oriented
closer to the line of sight. The observed trend of $\Delta PA$ with
emission line equivalent width (also noticed by \citealt{ASV96} in a
larger AGN sample) also supports this interpretation.

\subsubsection{Parsec-scale core dominance}
\begin{figure*}
\includegraphics[scale=1.0]{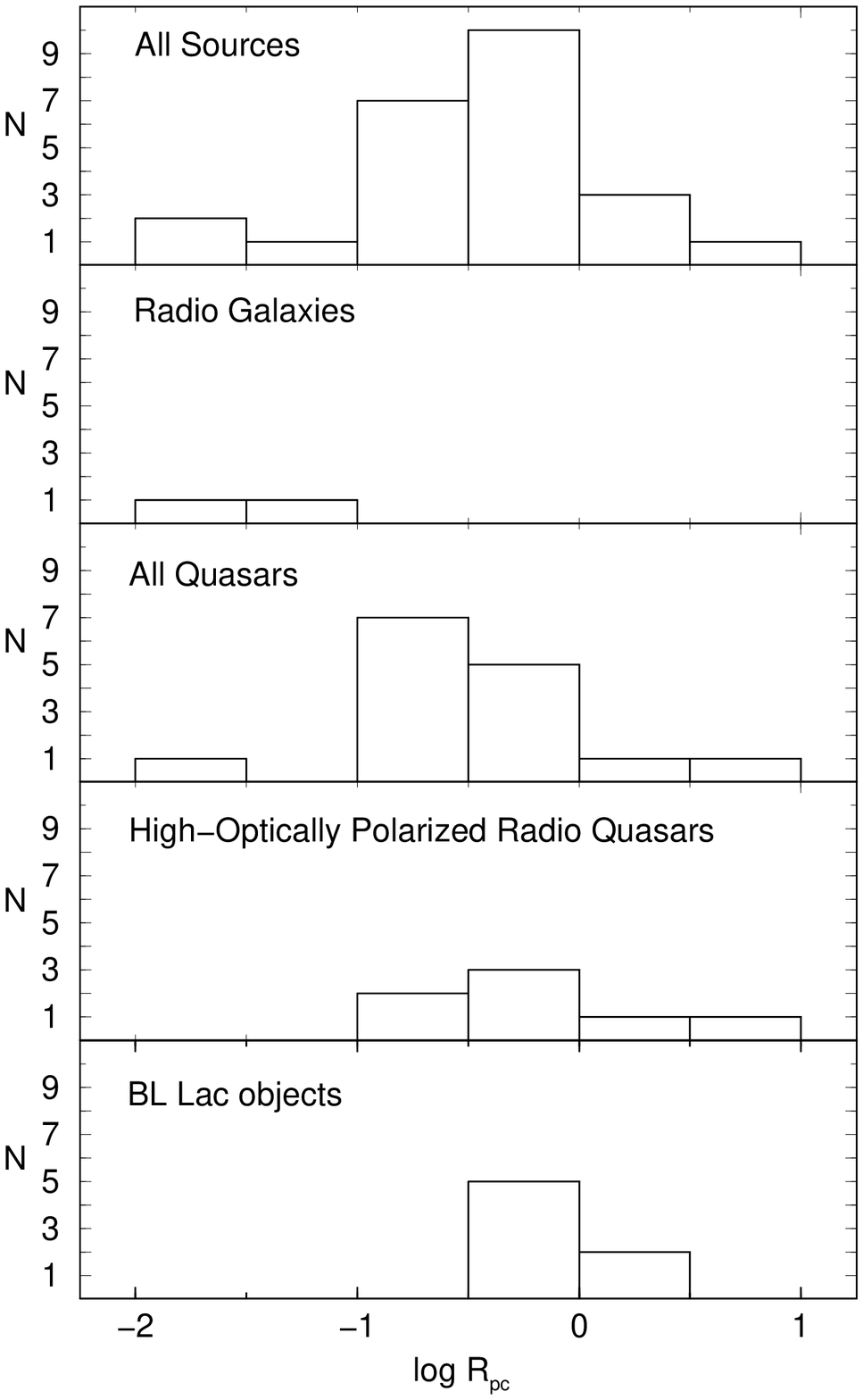}
%\plotone{f5.eps}
\caption{\label{R_pc_hist} Distribution of parsec-scale core dominance
($R_{pc}$) for various optical sub-classes of the FS-PR sample.}
\end{figure*}

Numerous studies \citep[e.g.,][]{ILT91,WWB92} have found that
parsec-scale core dominance ($R_{pc}$) is likely a good beaming
indicator, given that it is well-correlated with several other source
properties that are affected by beaming. These include a strong
anti-correlation with emission line equivalent width \citep{ILT91},
and a positive correlation with optical polarization \citep{ILT91},
both of which we confirm here (Table~\ref{corr_results}).  In
Figure~\ref{R_pc_hist} we show the distribution of parsec-scale core
dominance for the various optical classes in the FS-PR sample. Our K-S
tests reveal no statistically significant differences in these
distributions.

\subsubsection{Kiloparsec-scale core dominance}
\begin{figure*}
\plotone{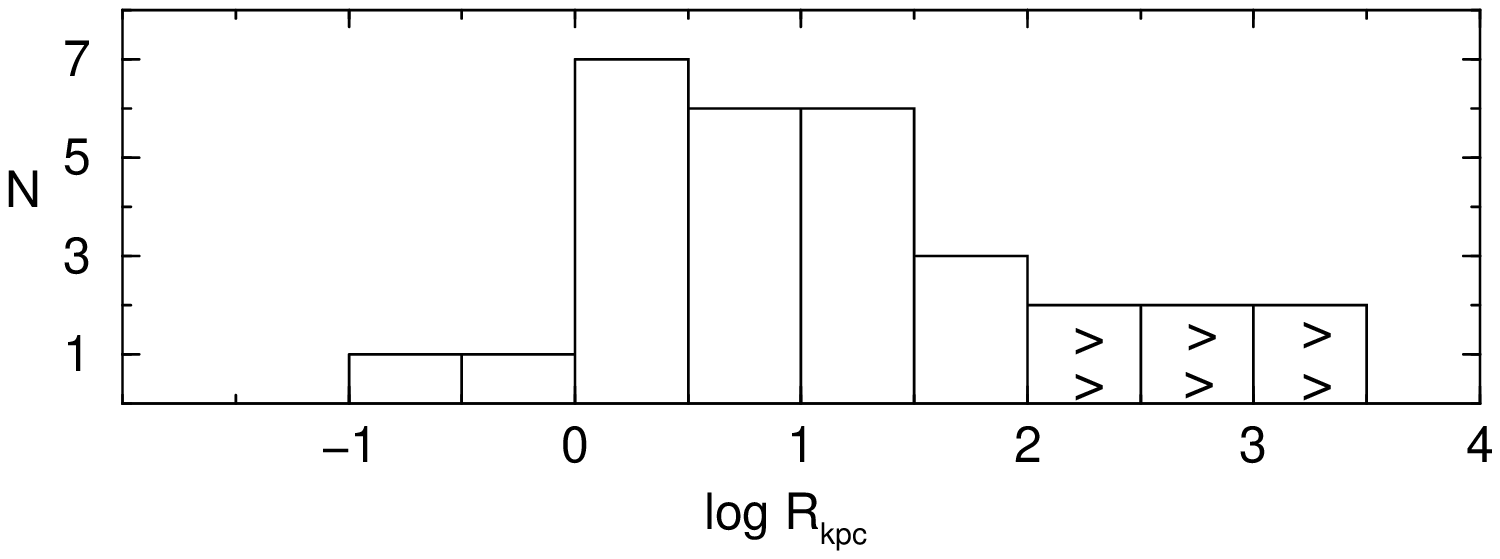}
\caption{\label{R_kpc_hist} Distribution of kiloparsec-scale core dominance
($R_{kpc}$) for the FS-PR sample. Sources with lower limits are
indicated by right arrows.}
\end{figure*}

In Figure~\ref{R_kpc_hist} we show the distribution of
kiloparsec-scale core dominance for the FS-PR. The measured values and
limits span a remarkable range of at least four orders of magnitude,
which reflects the wide variety of extended structure present in the
sample (the parsec-scale core dominance spans only $\sim2.5$ orders of
magnitude in comparison). We confirm the finding of \cite{MOF95} that
more highly core-dominated sources tend to have weaker extended
luminosities.

The core dominance on kiloparsec scales is not correlated with any
parsec-scale quantity for the FS-PR, such as emission line equivalent
width, or optical polarization, as is the case for the parsec-scale
core dominance. These correlations have been found, however, for
non-core selected samples, e.g., \cite{BM87}. It appears that VLA
core dominance is not a useful indicator of inner jet orientation for
samples selected on the basis of core flux density.

\subsubsection{Apparent jet velocity}
The maximum apparent jet velocity of a source is not necessarily a
good indicator of relativistic beaming and jet orientation in
core-selected samples, due to the non-linear dependence of apparent
velocity ($\beta_{app}$) on viewing angle. Since the apparent velocity
varies as $\beta\cos{\theta} / (1-\beta\cos{\theta})$, where $\beta$
is the true jet velocity in units of $c$ and $\theta$ is the viewing
angle, highly beamed jets aligned within $1/\Gamma = \sqrt{1-\beta^2}$
radians to the line of sight can have rather low apparent
velocities. We did not detect any correlations between apparent
velocity and any other source property for the 20 FS-PR sources for
which proper motion data are available. It appears that the full PR
sample, which contains a wider range of jet viewing angles, may be
better suited for exploring trends with apparent velocity.

\subsubsection{Gamma-ray emission}
The FS-PR sample contains seven AGNs that were detected in high-energy
gamma-rays by the EGRET observatory. Our two-sample tests showed no
significant differences in the properties of the EGRET versus
non-EGRET objects. It is possible, however, that this null result may
be due to the sample sizes (7 versus 25 objects), since the
statistical reliability of Kolmogorov-Smirnov tests is greatly
decreased for such small samples \citep{PTV92}.

\section{Discussion}
\subsection{\label{cluster}     Cluster analysis}

Our network diagram (Fig. \ref{network}) graphically displays the
relationships between the various properties of the FS-PR sources. In
order to investigate possible groupings in this diagram,
we have applied a cluster analysis to our data
\citep[e.g.,][]{N97}. This method searches for natural groupings among
variables that are most correlated. The most common (``single-link'')
method involves taking a symmetric matrix of correlation coefficients
$r_{ij}$ and finding the two variables $a$ and $b$ that are the best
correlated. These two variables are replaced by a new ``cluster''
variable $c$, for which $r_{ci} = \max[r_{ai},r_{bi}]$. The process is
repeated until there is only one cluster variable remaining. The
results are usually presented in the form of a tree diagram, in which
the clusters are represented by branches.

A few modifications to the standard cluster method were needed to apply
it to our data. Since our computed correlation coefficients for censored
variables have a different form of variance than the uncensored ones
\citep{AS96}, they cannot be directly inter-compared. Also, the
fact that many of our variables have a mutual dependence on redshift
will artificially skew the shape of the cluster tree diagram. We have
therefore taken the following steps in order to arrive at a more
accurate representation of the true cluster tree. First, we have used
the statistical significance levels in place of the correlation coefficients to
compare the variables. For the two nominal variables (IDV and extended
morphology class), we have used the two-sample test significance
levels. Second, we have used the significance levels with redshift
partialed out for all of the luminosity variables. Finally, we have
replaced the significance level for the rejected correlations listed
in Table~\ref{corr_results} with a value that excludes the outlier
point.

\begin{figure*}
\plotone{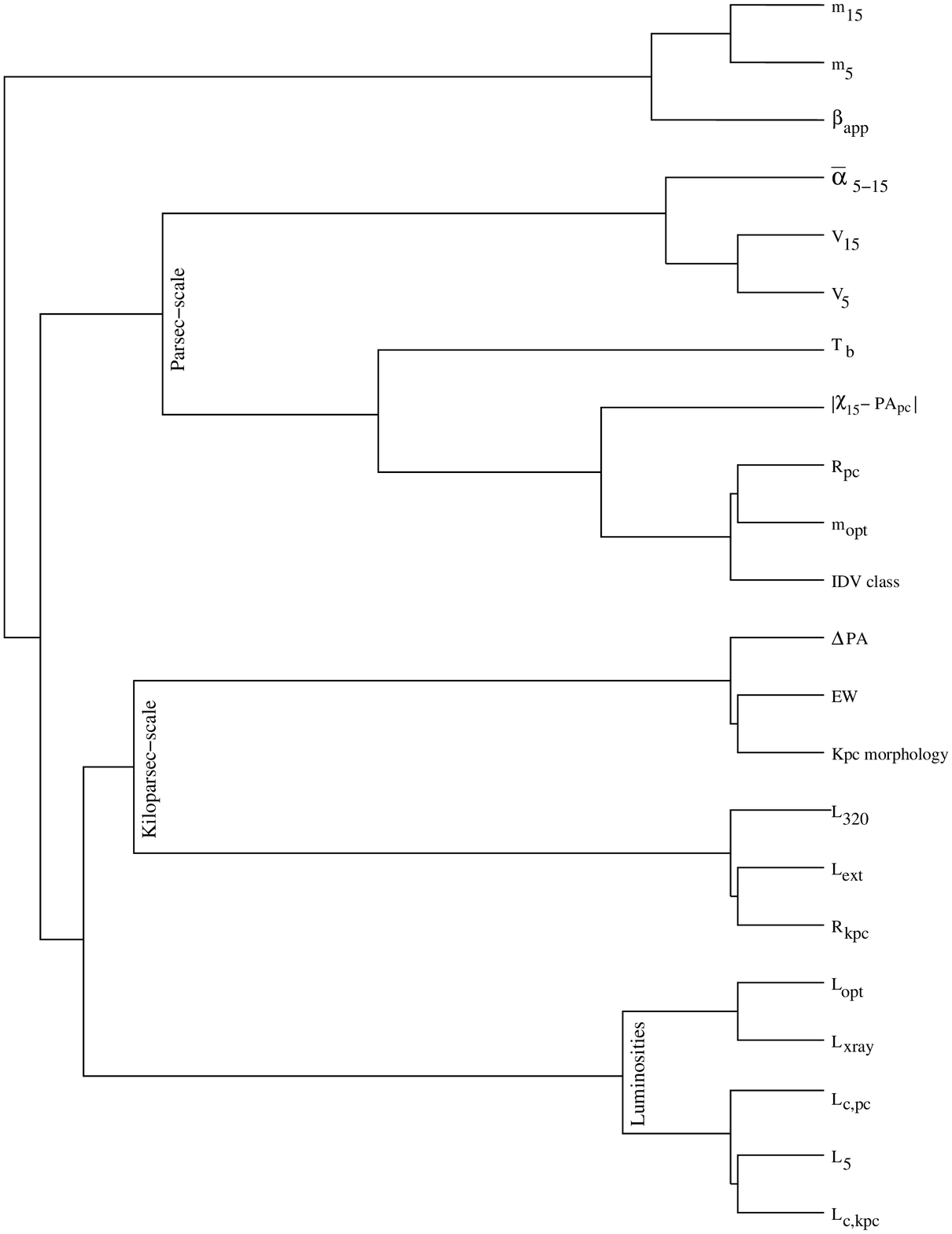}
\caption{\label{newtree} Cluster diagram for source properties of the
FS-PR sample. For symbol definitions, see Table~\ref{quantities}. }
\end{figure*}

We show the resulting tree diagram in Figure~\ref{newtree}. The
vertical connections indicate a close connection between variables and
groups of variables. The horizontal axis is roughly proportional to
correlation strength, with the vertical bars located furthest to the
right representing the strongest correlations. The luminosity
variables form a distinct grouping since they are poorly correlated
with the remaining variables. The apparent jet velocity and fractional
polarization levels at 5 \& 15 GHz also form a distinct group that is
largely independent of other source properties.

Among the remaining variables, those associated with the parsec-scale
core form a distinct grouping (labeled ``parsec scale'' in
Fig.~\ref{newtree}). The emission line equivalent width,
kiloparsec-scale morphology, and jet misalignment angle form a
separate grouping.  The exclusion of the equivalent width from the
parsec-scale grouping is somewhat unexpected, given that the broad
line region in quasars is thought to be less than a parsec in extent
\citep{CFP82}. This may imply that the kiloparsec-scale properties of
jets such as bending are influenced by the intrinsic line strength,
and in turn the density of the broad line gas.

\subsection{\label{BL_LAC}  Emission line strength and the BL Lacertae class}
Early VLBI studies of AGN samples with complete optical spectra
(e.g. \citealt{ILT91}) showed that emission line strength was
correlated with core dominance on parsec-scales, and led to a model in
which the optical continuum in highly beamed sources is dominated by
synchrotron emission from the jet \citep{WWB92}.  Since the emission
from the broad line region is unbeamed, the ratio of line to continuum
flux (as well as emission line equivalent width) should steadily
decrease with viewing angle to the source. Also, the overall optical
polarization should increase, as the polarized synchrotron component
begins to dominate. All of these predictions are borne out in the
FS-PR, as evidenced by the mutual correlations between equivalent
width, $\Delta PA$ and extended morphology.

These correlations have potentially important
implications for the BL Lacertae class of AGN, which by definition
have low emission line equivalent widths. Several studies have shown that
these objects have intrinsically different properties than their
broad-lined counterparts, the high optically-polarized quasars
(HPQs). These include differences in their parsec-scale magnetic
fields
\citep{CWRGB93}, X-ray spectral shapes \citep{WW90}, and extended
radio power \citep{PadM92}. It has been suggested that the parent
population of the BL Lacs are FR-I type radio galaxies, while the
radio quasars belong to the more powerful FR-II population (see
\citealt{UP95} and references therein).

The nine BL Lacs in the FS-PR rank in the top $30\%$ of the sample in
terms of optical polarization, $\Delta PA$ values, 5 GHz variability
amplitude, and parsec-scale core dominance.  This implies that they
are indeed highly beamed objects. There remains some debate over
whether BL Lacs are in fact more beamed (i.e., have higher Doppler
factors) than high-optically polarized quasars. The FS-PR sample is
not ideally suited to fully investigate this issue, since it contains
only nine BL Lacs and nine confirmed HPQs. Nevertheless, we have
performed several K-S tests on our data, and find that although the BL
Lacs do tend to have lower than average redshifts, there are otherwise
no significant differences in the $\Delta PA$, optical polarization,
variability amplitude, and core dominance distributions of the two
classes.  Therefore, despite strong indications that the observed
equivalent width in core-dominated radio sources is highly dependent
on beaming, there is insufficient evidence in the FS-PR to support a
claim that BL Lacs are more highly beamed than HPQs.  This
hypothesis should not be ruled out, however, until a similar analysis
can be conducted on a much larger sample such as the Caltech-Jodrell
Flat-spectrum survey (CJ-F; \citealt{TVRP96}).

\subsection{\label{simulations} Relativistic beaming simulations}
In the previous discussion we have made reference to general
predictions of the relativistic beaming model, and how it can account
for many of the correlations seen in the FS-PR sample. In this section
we describe in more detail the predicted properties of sources in
core-selected, flux-limited samples, based on Monte Carlo beaming
simulations.

Simulations of relativistic beaming by \cite{VC94} and others have
shown that core-selected samples should be dominated by objects with
jets pointing nearly directly at us. This would suggest that the FS-PR
should contain a narrow range of highly beamed, high-Doppler factor
sources. It is essential to consider, however, the additional effects
of the luminosity function (LF) and redshift distribution of the FS-PR
parent population. It is very likely that some low-Doppler factor
sources that happen to be intrinsically very luminous or relatively
nearby may have also been included in the FS-PR sample. This would
explain the presence of the two radio galaxies (3C 84 and 2021+614),
which are generally considered to be a weakly beamed class of AGN
\citep{UP95}. It would also account for the large range of kiloparsec-scale
core dominance found within the sample. The effects of adding a LF and
redshift distribution to beaming models have been investigated by
\cite{LM97}. They found that samples such as the FS-PR should in fact
contain a wide range of Doppler factors. In this section we use the
model of \cite{LM97} to show how many of the correlations in the FS-PR
are likely a result of a mutual dependence of source properties on
Doppler factor.

\cite{LM97} used their Monte Carlo model to successfully fit various 
properties of the Caltech-Jodrell Flat-spectrum sample (CJF;
\citealt{TVRP96}), of which the FS-PR is a sub-set. Their best-fit 
parent population contains a power law jet Lorentz factor distribution
of the form $N(\Gamma) \propto \Gamma^{-1.25}$ with $1.001252 < \Gamma
< 30$.  The jets are randomly oriented in space, and distributed with
a constant co-moving space density out to $z=4$. The jet luminosity
function mirrors that of the FR-II radio galaxies, which was derived
by \cite{UP95}, and incorporates pure exponential luminosity
evolution. The jets are assumed to be two-sided, and have luminosities
that are Doppler boosted by a factor of $\delta^2$. In a typical model
run, parent objects are continuously created until 293 sources with
flux density exceeding 0.35 Jy (the CJF sample size and cutoff) are
obtained. Further details of the model can be found in \cite{LM97}.

\begin{figure*}
\plotone{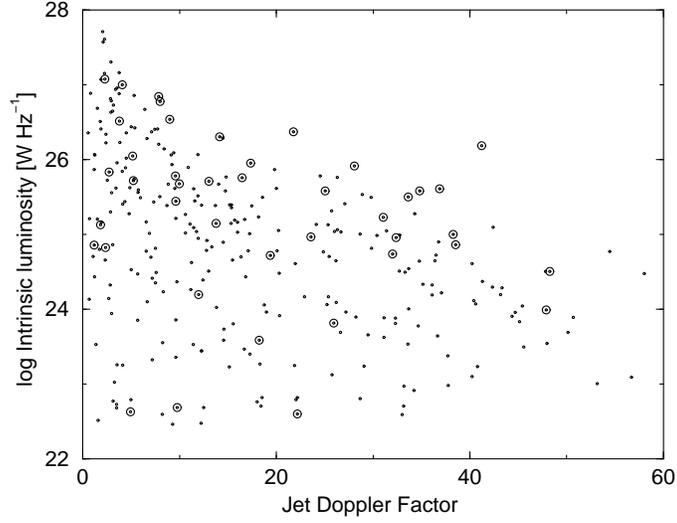}
\caption{\label{doppler__Lint} Plot of intrinsic (unbeamed) 5 GHz
jet luminosity vs. Doppler factor for a simulated Caltech-Jodrell
Flat-spectrum sample (small circles) and  a simulated FS-PR sample
(large circles). }
\end{figure*}

\begin{figure*}
\plotone{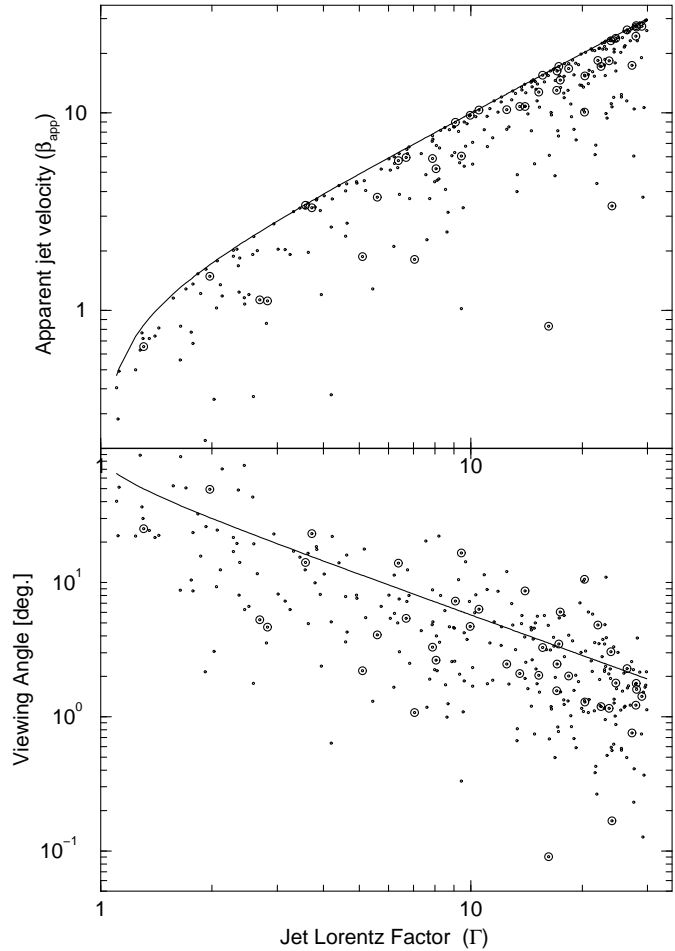}
\caption{\label{gamma__theta} Upper panel: Plot of apparent velocity
vs. Lorentz factor for a simulated Caltech-Jodrell
Flat-spectrum sample (small circles) and  a simulated FS-PR sample
(large circles). Bottom panel: Plot of viewing angle vs. Lorentz
factor. The solid lines in both panels indicate the position of
sources seen at a viewing angle that maximizes the apparent velocity.}
\end{figure*}

\begin{figure*}
\plotone{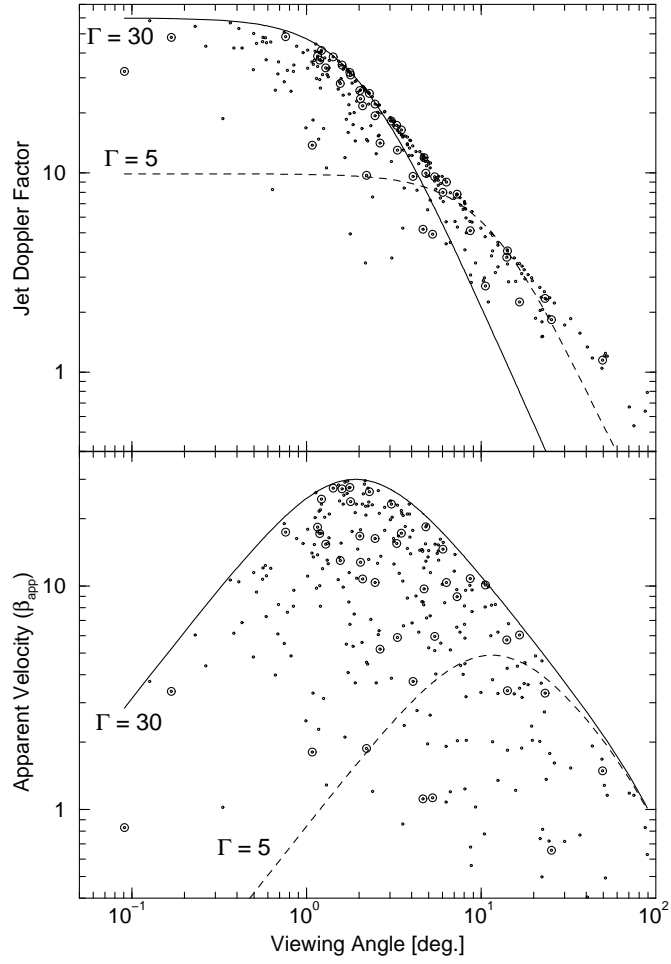}
\caption{\label{beta_theta} Upper panel: Plot of jet Doppler factor 
vs. viewing angle for a simulated Caltech-Jodrell Flat-spectrum sample
(small circles) and a simulated FS-PR sample (large circles). Bottom
panel: Plot of apparent velocity vs. viewing angle. The solid lines in
both panels indicate the position of a source with Lorentz factor 30,
and the dashed lines indicate a source with Lorentz factor 5.}
\end{figure*}

In Figures \ref{doppler__Lint}, \ref{gamma__theta}, and
\ref{beta_theta} we plot the results of a single Monte Carlo
simulation with the above parameters that illustrates several of the
strong biases that are present in flux-limited samples. The small
circles represent the 293 simulated CJF sources, and larger circles
indicated sources brighter than 1.3 Jy (the FS-PR cutoff).
Figure~\ref{doppler__Lint} plots the intrinsic luminosity of the jet
(removing the Doppler boosting effect) against Doppler factor. The
upper envelope to the distribution indicates that extremely powerful
sources that are also highly beamed are exceedingly rare, and that
weakly beamed sources generally need to be more intrinsically luminous
to be included in flux-limited samples. This trend provides a natural
explanation for the observation that IDV type 0 sources generally have
stronger 320 MHz and 5 GHz luminosities than the IDV type I and II
sources in the FS-PR, since the non-IDVs are expected to be less
relativistically boosted (see \S\ref{IDV}).

In Figure~\ref{gamma__theta} we plot two quantities against the jet
Lorentz factor: apparent velocity (upper panel), and viewing angle
(lower panel). The solid lines on these panels indicate the positions
of sources seen at the viewing angle that maximizes the
apparent velocity ($\theta = \arcsin{[1/\Gamma]}$). As pointed out by
\cite{LM97}, there is a substantial spread around this angle (lower
panel), with most sources seen at significantly smaller angles to the
line of sight. Nevertheless, the fairly good correlation in the upper
panel suggests that apparent velocity can be used as a statistical
indicator of the Lorentz factor. We note, however, that our
simulations assume the observed pattern speed reflects the true speed
of the jet, which may not be the case for all sources \citep{VC94}.

The apparent velocity of a jet by itself is {\it not} a good indicator
of viewing angle (lower panel of Fig.~\ref{beta_theta}), due to the
preponderance of jets seen within the $1/\Gamma$ viewing cone. At best
it can be used to set an upper limit on the viewing angle, through the
well known formula $\theta \le
\arccos{[(\beta_{app}^2-1)/(\beta_{app}^2+1)]}$. The lack of a good
correlation between apparent velocity and viewing angle may explain
why the former quantity is not correlated with other source properties
in the FS-PR. Alternatively, the measured apparent speeds could simply
be pattern speeds that do not reflect the true jet Lorentz factor. If
this is the case, the best statistical approach would be to use
long-term VLBI monitoring to determine a maximum observed velocity for
each source in a large sample.

The upper panel of Fig.~\ref{beta_theta} shows that we can expect an
extremely good correlation between viewing angle and Doppler factor in
a small flux-limited sample such as the FS-PR, provided the parent
population has an upper Lorentz factor limit. This trend is likely the
source of correlations between orientation-dependent properties (e.g.,
jet misalignment angle and arcsecond-scale morphology) and Doppler
boosting-dependent properties (e.g., core dominance and emission line
equivalent width).

\subsection{\label{recognition} The identification of highly beamed
radio sources}

The Doppler factors of extragalactic jets are notoriously difficult to
measure accurately, due to a lack of good indicators of true jet
velocity and orientation. Although some highly beamed sources can be
recognized on the basis of blazar properties (i.e., high variability
and optical polarization), other beamed sources may have intrinsically
low variability and/or polarization.  Our analysis of the FS-PR sample
suggests a set of criteria that can be used to identify the AGNs with
the highest Doppler factors in core-selected samples. In general, the
most highly beamed sources will tend to have a) flat radio spectra, b)
a type I or II IDV classification; c) one-sided, halo-like, or no
extended structure on kiloparsec scales; d) a relatively large
($\gtrsim 60\arcdeg$) misalignment between their jet directions on
parsec and kiloparsec scales; e) optical polarization that has
exceeded $\sim 3\%$ on at least one occasion; f) relatively high core
dominance on parsec ($R_{pc} \gtrsim 0.3$) and kiloparsec ($R_{kpc}
\gtrsim 10$) scales; g) an equivalent width of the largest permitted
emission line smaller than $\sim 15$ \AA \ (source frame); and h) a 5
GHz source frame core brightness temperature exceeding $10^{11} \rm
K$.

After taking into account the missing data on some of our sources,
there are three HPQs (0133+476, 0804+499, and 1739+522) and six BL
Lacs (0454+844, 0954+658, MK 501, 1749+701, 1803+784, and BL Lacertae)
in the FS-PR that are consistent with the above criteria.  We suggest
that these sources have significantly higher Doppler factors than the
other sources in the FS-PR, and should be considered as prime targets
for high-frequency space-VLBI and IDV monitoring observations. We
predict that two of these objects (0454+844 and 1739+522) will
eventually be found to display type I or II IDV, based on their other
properties. 

\section{Conclusions and suggestions for future work}

We have obtained data using the facilities of the VSOP mission and
from the literature on a flat-spectrum, core-selected subsample of the
Pearson-Readhead AGN survey (the FS-PR sample) in order to investigate
the effects of relativistic beaming in compact radio sources. The main
results of our multi-dimensional correlation analysis on these data
are as follows:

(i) The properties of sources that have displayed type I or II IDV are
significantly different than those that have shown no IDV
activity. They have flatter spectral indices, higher core
dominance  on parsec scales, higher optical polarizations and smaller
emission line widths. They also have lower 320 MHz and 5 GHz
luminosities and weaker kiloparsec-scale core components. These
findings are all consistent with Monte Carlo beaming simulations
which predict that higher Doppler factor sources in flux-limited
generally have weaker intrinsic luminosities. 

(ii) The observed luminosity properties of the FS-PR are generally
well-correlated with each other, but not with other properties of the
sample. The only exception is the trend with IDV described above.

(iii) The high-optically polarized sources in the FS-PR show a trend
of increased parsec-to-kiloparsec jet misalignment ($\Delta PA$) with
decreasing optical polarization. However, the lowest optically
polarized sources ($m_{opt} < 3\%$) generally have small jet
misalignments, which may suggest that they have intrinsically
different jet properties.

(iv) We find that sources with two-sided radio morphology on
kiloparsec-scales have significantly different properties than other
sources in the FS-PR.  They tend to be less core dominated on parsec-
and kiloparsec scales, and have stronger extended luminosities and
larger emission line widths. Their jets are also less misaligned
between parsec- and kiloparsec scales, which suggests that they are
oriented at angles farther away from the line of sight.

(v) We have used high-resolution VLBI images to confirm the
bi-modality of the $\Delta PA$ distribution for the PR sample
\citep{PR88}, with a division occurring $\Delta PA \simeq
60\arcdeg$. The aligned sources tend to have broader emission line
equivalent widths than the misaligned population.

(vi) We did not find any correlations between maximum apparent jet velocity
and other source properties. This may be due to the predicted weak
dependence of $\beta_{app}$ on viewing angle for core-selected
samples, or the fact that the measured speeds may not reflect the true
bulk speeds of the emitting material in the jet. 

(vii) The nine BL Lacertae objects in the FS-PR rank among the top
$30\%$ of the sample in terms of their optical polarization, $\Delta
PA$ values, 5 GHz variability amplitude, and parsec-scale core
dominance, which suggests that they are highly beamed objects. We
found no significant differences in their properties as compared to the
high-polarization quasars, other than the fact that they have smaller
redshifts.

(viii) Our correlation analysis and Monte Carlo beaming simulations
have shown that the properties of the FS-PR sample are remarkably
consistent with the predictions of the relativistic beaming model. The
majority of the observed correlations are likely due to a mutual
dependence of source properties on the jet Doppler factor and viewing
angle, and do not necessarily reflect intrinsic relations among jet
properties. 

Our work has demonstrated the merits of obtaining a wide variety
of data on a relatively small sample, and suggests several further
observations that can improve our understanding of the beaming
phenomenon in AGNs. These include optical variability monitoring of
the PR survey to further investigate the connection between optical
and radio properties, and more uniform monitoring of the sample for
IDV activity in the radio. It will also be important to measure
apparent velocities for the remaining sources in the sample, and to
obtain high-dynamic range VLA and MERLIN maps down to a uniform
luminosity level. The latter can be used to trace out bending in the
jets as they propagate from parsec- to kiloparsec scales, and examine
how this phenomenon is influenced by other source properties. With
regular monitoring the optical polarization of the sample at regular
intervals, it should be possible to establish a duty cycle of optical
polarization in blazars, and to better understand the inherent
differences in high- and low-optically polarized AGNs.

%-------------------------------------------------------------------
% Acknowledgments
%----------------------------------------------------------------------
\acknowledgments
 We thank D. W. Murphy for useful discussions during the preparation
 of this manuscript. This research was performed in part at the Jet
 Propulsion Laboratory, California Institute of Technology, under
 contract to NASA, and has made use of data from the following
 sources:

 The NASA/IPAC Extragalactic Database (NED), which is operated by the
 Jet Propulsion Laboratory, California Institute of Technology, under
 contract with the National Aeronautics and Space Administration.

 The University of Michigan Radio Astronomy Observatory, which
 is supported by the National Science Foundation and by funds from the
 University of Michigan.

% Tables
\begin{deluxetable}{llll}
%\tabletypesize{\scriptsize}
\tablecolumns{4}
\tablecaption{\label{quantities}Measured and Derived Quantities for the Pearson-Readhead AGN Sample}
\tablewidth{0pt}
\tablehead{\colhead{Symbol} & \colhead{Property} &\colhead{Units} & \colhead{Reference}}
\startdata
\sidehead{General Properties}
\n &     EGRET gamma-ray detection/non-detection & \n &  \citealt{HBB99}  \\
\n &	Optical classification (RG, LPRQ, HPQ, BL Lac) & \n &Various \\
\n & Kiloparsec-scale extended morphology & \n &Various \\
$z$ & Redshift &\n & \citealt{LZR96} \\
$EW$ & Equivalent width of widest permitted line (source frame) &
\AA &  \citealt{LZR96} \\
%$L/C$ & Ratio of optical line to continuum emission &\n & \citealt{ILT91}  \\
$\bar \alpha_{5-15}$  &   Time-averaged spectral index between 4.8 and 14.5 GHz &\n & \citealt{AAH92} \\

\sidehead{Variability Properties}
$V_5$ & Variability amplitude at 5 GHz &\n & \citealt{AAH92} \\
$V_{15}$ & Variability amplitude at 14.5 GHz &\n & \citealt{AAH92} \\
%$T_{b,var}$ & Variability brightness temperature &   & \citealt{LV99}\\
\n &  IDV classification & \n &  Various \\

\sidehead{Luminosity Properties} 
$L_{320}$ &  Total luminosity at 320 MHz & $\rm W\; Hz^{-1}$ &  \citealt{ILT91} \\
$L_{5}$ &  Total luminosity at 5 GHz &$\rm W\; Hz^{-1}$ & UMRAO database\tablenotemark{a} \\
$L_{opt}$ & Optical V band luminosity  & $\rm W\; Hz^{-1}$ & NED\footnote{http://nedwww.ipac.caltech.edu/} \\
$L_{xray}$ &  X-ray luminosity at 1 keV & $\rm W\; Hz^{-1}$ &Various\\
$L_{c,pc}$ & Parsec-scale core luminosity at 5 GHz &  $\rm W\; Hz^{-1}$ &This paper \\
$L_{c,kpc} $ & Kiloparsec-scale core luminosity at 1.4 GHz &  $\rm W\; Hz^{-1}$ & Various\\
$L_{ext} $ & Total arcsecond-scale extended luminosity at 1.4 GHz &  $\rm W\; Hz^{-1}$ &Various \\

\sidehead{Jet Properties} 
$T_b$   & Brightness temperature of VLBI core component & K &
\cite{TLP00} \\ 
$R_{pc}$ & Ratio of parsec-scale core to remaining flux density & \n & This paper \\
$R_{kpc}$ & Ratio of kiloparsec-scale core to remaining flux density & \n & Various \\
$\beta_{app}$ & Fastest measured apparent speed of jet component & \n &Various \\ 
$\Delta {PA}$ & Difference in position angle between pc-and kpc-scale
jet & Deg. & Various \\ 

\sidehead{Polarization Properties}
$m_{opt}$ &	Optical V band linear polarization & $\%$ & \citealt{ILT91} \\
$m_5$ &   Mean linear polarization at 4.8 GHz  & $\%$ & \citealt{AAH92} \\
$m_{15}$ &   Mean linear polarization at 14.5 GHz &  $\%$ &\citealt{AAH92} \\
$|\chi_{15} - PA_{pc}|$ & Offset of integrated 14.5 GHz $\chi$
w.r.t. pc-scale jet PA & Deg. &  \citealt{AAH92} \\

\enddata 
\tablenotetext{a}{http://www.astro.lsa.umich.edu/obs/radiotel/umrao.html}
\tablecomments{Data on properties marked ``Various'' were gathered
from multiple sources (see \S3).} 
 \end{deluxetable}
\begin{deluxetable}{lllccrrrrrrrrcrc}
\tabletypesize{\scriptsize}
\rotate
\tablecolumns{16}
\tablecaption{\label{genprops}General Properties of Pearson-Readhead Sources}
\tablewidth{0pt}
\tablehead{ \colhead{IAU} &\colhead{Other}  &\colhead{Opt.} & &\colhead{IDV} &
 &     &\colhead{} &\colhead{} &\colhead{log}
&  & &\colhead{log} & & & \\
\colhead{Name} &\colhead{Name}& \colhead{Type} &\colhead{Morph.} & \colhead{Class} &  \colhead{z}
 &  \colhead{EW}       &\colhead{$L_{c,pc}$} &\colhead{$L_5$} &  \colhead{$R_{pc}$}
  &\colhead{$L_{c,kpc}$} & \colhead{$L_{ext}$} &  \colhead{$R_{kpc}$} & \colhead{Ref.} & \colhead{$L_{xray}$}& \colhead{Ref.} \\ 
%& & & & & & \colhead{[$\AA$]} & & & \colhead{[$\rm W\; Hz^{-1}$]}& \colhead{[$\rm W\; Hz^{-1}$]}
%& \colhead{[$\rm W\; Hz^{-1}$]}& \colhead{[$\rm W\; Hz^{-1}$]}\\
\colhead{ [1]} & \colhead{ [2]} & \colhead{ [3]}  &
\colhead{ [4]} & \colhead{ [5]} & \colhead{ [6]} &
\colhead{ [7]}& \colhead{ [8]}  & \colhead{ [9]} &\colhead{ [10]}& \colhead{ [11]} 
 & \colhead{ [12]} & \colhead{ [13]}  & \colhead{ [14]} & \colhead{ [15]} & \colhead{ [16]} }
\startdata
\sidehead{FS-PR sample objects}
 0016+731  &  \n    & Q&   2&   I&    1.781&     17.0&    26.99&    27.93& --0.89&    27.90& $<$25.57&  $>$2.33&        1  &     20.97 &   8 \\ 
 0133+476  &  OC 457   & HPQ&   C&  II&    0.859&     10.4&    27.46&    27.53& 0.75&    27.34& $<$24.26&  $>$3.08&        1  &     20.96 &   8 \\ 
 0212+735  &  \n    & HPQ&   C&   0&    2.367&     15.0&    28.25&    28.78& --0.37&    28.56& $<$25.59&  $>$2.97&        1  &     22.01 &   8 \\ 
 0316+413  &  3C 84    & RG &   2&   0&    0.017&    417.5&    24.07&    25.22& --1.12&    25.05&    24.58&     0.47&      2  &     18.00 &   9 \\ 
 0454+844  &  \n    & BL &   C&  \n&    0.112&   $<$1.0&       \n&    25.30&  \n&    25.09& $<$22.71&  $>$2.38&        3  &     18.01 &  10 \\ 
 0723+679  &  3C 179   & LPRQ&   2&  \n&    0.844&     39.2&       \n&       \n&  \n&    26.81&    27.76&   --0.95&        4  &        \n &  \n \\ 
 0804+499  &  OJ 508   & HPQ&   1&  II&    1.432&     13.0&       \n&    27.90&  \n&    27.53&    25.75&     1.78&        3  &     21.23 &   8 \\ 
 0814+425  &  OJ 425   & BL &   1&  II&    0.245&      1.0&    25.83&    26.19& --0.11&    26.38&    25.07&     1.30&      3  &     18.95 &  10 \\ 
 0836+710\tablenotemark{a}  &  4C 71.07 & Q&   2&  0&    2.180&   17.6&  27.88&    28.51& --0.52&    28.67&   27.51& 1.16&  3  &  22.83 &   8 \\ 
 0850+581  &  4C 58.17 & LPRQ&   2&  \n&    1.322&     25.9&       \n&    27.69&  \n&    27.48&    26.73&     0.75&        4  &     20.79 &  11 \\ 
 0859+470  &  4C 47.29 & Q&   2&   0&    1.462&     30.0&    27.39&    27.95& --0.42&    27.96&    27.55&     0.40&     3  &     20.70 &  11 \\ 
 0906+430  &  3C 216   & HPQ&   2&   I&    0.670&      3.1&    26.74&    27.32& --0.45&    26.92&    27.22&   --0.30&   5  &     20.23 &   8 \\ 
 0923+392  &  4C 39.25 & LPRQ&   2&   0&    0.699&     58.3&    26.52&    28.11& --1.58&    27.52&    27.09&     0.43&    3  &     20.99 &   8 \\ 
 0945+408  &  4C 40.24 & LPRQ&   1&   I&    1.252&     15.9&       \n&    27.78&  \n&    27.69&    26.86&     0.83&        3  &     20.93 &   8 \\ 
 0954+556\tablenotemark{a}  &  4C 55.17 & HPQ&   2&  \n&    0.900&      7.1&  \n&       \n&  \n&    27.72&    27.11&   0.61&  3  &   20.53 &   8 \\ 
 0954+658\tablenotemark{a}  &  \n    & BL &   1&  II&    0.367&      1.9&      \n&    26.48&  \n&    26.23&    25.17&   1.06&  6  &  19.84 &  10 \\ 
 1624+416  &  4C 41.32 & Q&   1&   0&    2.550&     13.6&    27.42&    28.39& --0.92&    28.45& $<$25.85&  $>$2.60&        3  &        \n &  \n \\ 
 1633+382\tablenotemark{a}  &  4C 38.41 & HPQ&   2&   0&    1.807&     38.1&  27.58&  28.36& --0.70&    28.27&    26.79& 1.47&  3  & 21.88 &   8 \\ 
 1637+574  &  OS 562   & Q&   H&  \n&    0.749&     17.3&    26.53&    27.01& --0.30&    27.22&    26.52&     0.70& 3  &     20.62 &  11 \\ 
 1641+399  &  3C 345   & HPQ&   2&   0&    0.595&     34.9&    26.86&    27.86& --0.96&    27.88&    27.11&     0.76& 3  &     20.69 &   8 \\ 
 1642+690  &  4C 69.21 & HPQ&   2&  II&    0.751&     26.0&    26.75&    27.22& --0.29&    27.15&    26.87&     0.29&3  &        \n &  \n \\ 
 1652+398  &  MK 501   & BL &   H&   I&    0.033&      0.8&    24.09&    24.66& --0.44&    24.57&    23.27&     1.29&5  &     19.37 &  10 \\ 
 1739+522\tablenotemark{a}  &  4C 51.37 & HPQ&   1&  \n&    1.381&      8.1&    27.93&    28.07& 0.43&    28.02& 26.27& 1.75&  4  & 21.20 &  12 \\ 
 1749+701  &  \n    & BL &   1&  II&    0.770&      0.2&    26.60& 26.98& --0.15&       \n&       \n&     \n &   14      &     20.55 &  10 \\ 
 1803+784  &  \n    & BL &   1&  II&    0.680&      2.9&    27.20&    27.38& 0.29&    27.26&    26.08&     1.18&        3  &     20.61 &  10 \\ 
 1807+698  &  3C 371   & BL &   2&   I&    0.050&      6.0&    24.39&    25.01& --0.50&    24.89&    24.81&     0.07& 5  &     18.32 &  10 \\ 
 1823+568  &  4C 56.27 & BL &   1&   I&    0.663&      1.1&    26.94& 27.23& 0.02&    26.88&    26.88&     0.00&        3  &     20.84 &  10 \\ 
 1928+738  &  4C 73.18 & LPRQ&   2&   I&    0.302&    398.9&    25.98&    26.92& --0.89&    26.83&    25.84&     0.99&  3  &     20.52 &   8 \\ 
 1954+513  &  OV 591   & Q&   2&  \n&    1.223&     18.3&    27.08&    27.79& --0.61&    27.72&    27.61&     0.11&  7  &        \n &  \n \\ 
 2021+614  &  OW 637   & RG &   C&   0&    0.228&     35.4&    25.01&    26.56& --1.54&    26.45& $<$23.19&  $>$3.27& 1  &        \n &  \n \\ 
 2200+420\tablenotemark{a}  &  BL Lac   & BL &   H&   I&    0.069&      6.8& 25.27&    25.72& --0.27&    25.57&  23.70& 1.87&  5  & 19.04 &  10 \\ 
 2351+456\tablenotemark{a}  &  4C 45.51 & Q&   2&   0&    1.986&     25.3&  \n&    28.16&  \n&    \n& \n &  \n &        13  &        \n &  \n \\ 
\sidehead{VSOP-PR sample objects}
  0153+744  &  \n    & Q&   C&   I&    2.338&     14.6&    27.71&    28.96& --1.23&    28.41& $<$25.53&  $>$2.88&        1  &     22.36 &   8 \\ 
 0711+356  &  OI 318   & Q&   2&  II&    1.620&     19.1&    26.72&    28.20& --1.47&    28.02&    25.79&     2.23&  3  &     20.77 &  \n \\ 
 1828+487  &  3C 380   & LPRQ&   2&  \n&    0.692&     36.1&    26.79&    27.91& --1.09&    27.73&    28.15&   --0.42&  3  &     21.08 &   9 \\ 

\enddata 

%\tablenotetext{a}{Measured at 5 GHz \citep{KWRG92} }
\tablenotetext{a}{Member of third EGRET catalog \citep{HBB99}}
\tablecomments{All quantities are calculated assuming $h = 0.65$, $q_o
= 0.1$, and $\Lambda = 0$. Luminosities are given in $\rm W\; Hz^{-1}$.
Columns are as follows: (1) IAU source name; 
 (2) Alternate name; 
 (3) Optical classification  (see \S\ref{gendesc}); 
 (4)  Kiloparsec-scale radio morphology, where C = compact, 1 = one-sided,
     2 = two-sided, and H = halo; 
 (5)  IDV classification (see \S\ref{variabdesc}); 
 (6)  Redshift; 
 (7)  Source frame equivalent width of widest permitted line [\AA];
 (8)  Luminosity of parsec-scale core component at 5 GHz; 
 (9)  Total (single-dish) luminosity of source at 5 GHz; 
 (10)  Ratio  of parsec-scale core flux density to remaining (non-core) flux density at 5 GHz
    in source rest frame; 
 (11)  Luminosity of kiloparsec-scale core component at 1.4 GHz; 
 (12)  Luminosity of kiloparsec-scale extended structure at 1.4 GHz; 
 (13)  Ratio of kiloparsec-scale core to extended flux density at
    1.4 GHz, in source rest frame; 
 (14)  Reference for columns 4, 11 and 12; 
 (15)  X-ray luminosity at 0.1 keV; 
 (16)  Reference for X-ray data.}

\tablerefs{\scriptsize (1)  \citealt*{XRP95}; (2) \citealt*{PGD90}; 
(3) \citealt*{MBP93}; (4) \citealt*{RSA95}; (5) \citealt*{AU85}; 
(6) \citealt*{CSB99}; (7) \citealt*{BP86};
(8) \citealt*{S97}; (9) \citealt*{HW99}; (10) \citealt*{USW96};
(11) \citealt*{PPG98}; (12)  \citealt*{CFG97}; (13) \citealt*{NRH95};
(14) \citealt*{OBC88}.}

\end{deluxetable}

\begin{deluxetable}{lrrrrrl}
\tabletypesize{\scriptsize}

\tablecolumns{7}
\tablecaption{\label{jetprops} Jet Properties of Pearson-Readhead Sources}
\tablewidth{0pt}
\tablehead{ \colhead{Source} &\colhead{$\beta_{app}$}     &
 &  \colhead{ $PA_{pc}$}  &\colhead{ $PA_{kpc}$} 
&\colhead{$\Delta {PA}$}  \\
\colhead{Name} & \colhead{$(v/c)$} &\colhead{Ref.} 
 &  \colhead{(deg)}    &  \colhead{(deg)}  &  \colhead{(deg)} &\colhead{Ref.} \\
\colhead{ [1]} & \colhead{ [2]} & \colhead{ [3]}  &
\colhead{ [4]} & \colhead{ [5]} & \colhead{ [6]} &
\colhead{ [7]}   }
\startdata
\sidehead{FS-PR sample objects}
0016+731&       17.3&  2 &        132&    169&  37& 22, 25 \\
0133+476&            \n &\n  &    330&     \n&  \n& 22,\n \\
0212+735&          8.6&   4&    121&     \n&  \n& 22,\n  \\
0316+413&          0.08&   5&     176&    162&  14& 22, 26  \\
0454+844&          1.1&   6&    177&     \n&  \n& 22,\n   \\
0723+679&          8.6&   8&      256&    266&  10& 22, 38 \\
0804+499&           \n&  \n&      127&    201&  74& 22, 27     \\
0814+425&           \n&  \n&     103&     50&  53& 22, 27    \\
0836+710&         21.7& 9&       201&    198&   3& 22, 29 \\
0850+581&          7.6&  10&      227&    151&  76& 22, 30 \\
0859+470&            \n&  \n&    357&    335&  22& 22, 28        \\
0906+430&          5.3& 11 &      151&    244&  93& 22, 31 \\
0923+392&          2.3& 12&        93&     76&  17& 21, 32 \\
0945+408&           \n&  \n&      137&     32& 105& 22, 28        \\
0954+556&           \n&  \n&      191&    301& 110& 23, 28       \\
0954+658&         12.2& 13&      326&    224& 102& 22, 33 \\
1624+416&           \n& \n&      261&    351&  90& \phn 1, 34       \\
1633+382&         15.9& 14&      279&    176& 103& 21, 27 \\
1637+574&         \n&  \n&         200&     \n&  \n& 22,\n               \\
1641+399&         29.3& 15&      284&    328&  44& 21, 35 \\
1642+690&         14.0&  16&      158&    186&  28&  22, 34 \\
1652+398&        5.6&    17&    160&     45& 115& 22, 36  \\
1739+522&           \n&  \n&      204&    260&  56&  22, 28       \\
1749+701&         10.9& 2&       275&     21& 106& 22, 25 \\
1803+784&          3.2& 13&      291&    193& 98& 22, 27 \\
1807+698&           \n&  \n&     252&    261&   9& 22, 36       \\
1823+568&         21.2&  14 &    196&    182&  14& 22, 34 \\
1928+738&         11.5&  4 &     166&    190&  24& 21, 27 \\
1954+513&           \n&   \n &  307&    338&  31& 22, 24 \\
2021+614&          0.19&  19 &   205&     \n&  \n& 19,\n     \\
2200+420&       13.0&  20 &     209&     \n&  \n& 22,\n      \\
2351+456&           \n&  \n&      321&    348&  27& 22, 39       \\

\sidehead{VSOP-PR sample objects}
0153+744&          3.8&   3&       68&     \n&  \n& 22,\n \\
0711+356&        $ <0.81$&  7&  329&    294&  35& 22, 27 \\
1828+487&         10.6&  18&     311&    311&  0 & 22, 37 \\

\enddata 

\tablecomments{\scriptsize Columns are as follows: (1) IAU source name; (2)
Maximum component speed in units of $c$,
assuming $h = 0.65$ and $q_o = 0.1$; (3) Reference for apparent
speed; (4) Parsec-scale jet position
angle; (5) Kiloparsec-scale jet position angle. (6) Difference in
parsec- and kiloparsec-scale jet position angles; (7) References for
images used in columns 4 and 5, respectively.}
\tablerefs{\scriptsize (1) \citealt*{L-PRI}; (2) \citealt*{SWH92}; (3) \citealt*{HKW97};
(4) \citealt*{WSJ88}; (5) \citealt*{DKR98}; (6) \citealt*{GMCW94}; (7)
\citealt*{CMP94}; (8) \citealt*{Por87}; (9) \citealt*{MMMH00}; (10)
\citealt*{BPR86}; (11) \citealt*{PFF00}; (12) \citealt*{FEK97}; (13)
\citealt*{GC96}; (14) S. G. Jorstad,  private communication;
(15) \citealt*{RZL00}; (16) \citealt*{PBL86}; (17) \citealt*{GCF00};
(18) \citealt*{PW98}; (19) \citealt*{TSS99}; (20) \citealt*{DMM00};
(21) \citealt*{LS00}; (22) Lister et al., in preparation; (23)
A. P. Marscher, private communication; (24) \citealt{KWR90}; (25)
\citealt*{XRP95}; (26) \citealt*{PGD90}; (27) \citealt*{MBP93}; (28)
\citealt*{RSA95}; (29) \citealt*{HMK92}; (30) \citealt*{SPZ85}; (31)
\citealt*{TGO95}; (32) \citealt*{MZS91}; (33) \citealt*{SDD91}; (34)
\citealt*{OBC88}; (35) \citealt{KWR89}; (36) \citealt*{CSB99}; (37)
\citealt*{LGS98}; (38) \citealt*{Aku92}; (39) \citealt{NRH95}.}
\end{deluxetable}

\begin{deluxetable}{llrrlrll}
%\tabletypesize{\scriptsize}
\tablecolumns{8}
\tablecaption{\label{corr_results}Correlations Among Observed Properties of the FS-PR Sample}
\tablewidth{0pt}
\tablehead{\colhead{Property} & \colhead{Property} & \colhead{N} & \colhead{$\tau$} & \colhead{P}
& \colhead{$\tau_z$} & \colhead{$P_z$} & \colhead{Notes} \\
\colhead{(1)} & \colhead{(2)} &\colhead{(3)} & \colhead{(4)} &\colhead{(5)} & \colhead{(6)} &\colhead{(7)} & \colhead{(8)} }
\startdata
%\sidehead{General Properties}
$L_{ext} $    &  $R_{kpc}$        &   30 &  --0.56 &$9.7\times 10^{-4}$&  --0.60 &$4.9\times 10^{-12}$& Expected  \\ 
$R_{pc}$   &  $EW$                &   24 &  --0.49 &$8.1\times 10^{-2}$&  --0.50 &$1.8\times 10^{-7}$&  \cite{ILT91} \\ 
$L_5$      &  $L_{c,kpc}$         &   28 &    0.91 &$1.1\times 10^{-9}$ &   0.77 &$7.6\times 10^{-7}$& \cite{AS96}  \\  
$V_{15}$          &  $V_5$        &   31 &    0.62 &$9.6\times 10^{-5}$&    0.63 &$7.2\times 10^{-5}$& \cite{AAH92}  \\  
$L_{ext} $    &  $L_{320}$        &   28 &    0.48 &$2.4\times 10^{-2}$&    0.49 &$9.2\times 10^{-4}$&  Expected \\  
$m_{opt}$          &  $EW$        &   30 &  --0.42 &$1.2\times 10^{-1}$ & --0.40 &$2.1\times 10^{-2}$& \cite{ILT91}  \\
$L_5$      &  $L_{xray}$          &   25 &    0.79 &$3.6\times 10^{-6}$ &   0.52 &$2.6\times 10^{-2} $&  \cite{BM87} \\  
$L_{opt}$      &  $L_{xray}$      &   26 &    0.67 &$1.4\times 10^{-4}$&    0.52 &$2.6\times 10^{-2}$ & \cite{FTE84} \\  
$R_{pc}$   &  $m_{opt}$           &   23 &    0.50 &$8.2\times 10^{-2}$&    0.52 &$4.5\times 10^{-2}$&   \cite{ILT91} \\  
$  \Delta {PA}$  &   $EW$  &          26 &  --0.39& $5.5\times 10^{-1}$ & --0.40 &$2.0\times 10^{-1}$ & \cite{ASV96} \\
$L_5$       &  $L_{320}$          &   27 &    0.69 &$5.1\times 10^{-5}$&    0.42 &$2.2\times 10^{-1}$&  \cite{ZB95} \\  
$L_5$ & $L_{c,pc}$                &   24 &    0.75 &$2.5\times 10^{-5}$&    0.41 &$4.8\times 10^{-1}$& \n  \\  
$L_{c,pc}$  &  $L_{c,kpc}$        &   23 &    0.76 &$4.6\times 10^{-5}$&    0.42 &$4.9\times 10^{-1}$& Expected  \\  
$m_{15}$            & $m_5$       &   31 &    0.34 &$7.2\times 10^{-1}$&    0.34 &$7.1\times 10^{-1}$& \cite{AAH92}  \\  
$L_{ext} $    &  $L_5$         &    28 &    0.33 & 1.1              &     0.25 &$ 9.3\times10^{-1}$ & Expected  \\
$L_{c,kpc}$    &  $L_{xray}$      &   25 &    0.71 &$2.5\times 10^{-5}$ &   0.41 &  1.2 & \cite{BM87} \\
$L_{c,pc}$   &  $L_{xray}$        &   20 &    0.72 &$1.0\times 10^{-3}$&    0.39 &  1.6    & \cite{BMM99}  \\  
$L_{opt}$    &  $L_{c,kpc}$  &        30 &    0.60 &$3.2\times 10^{-4}$ &   0.31 &  1.7 & \cite{FGG84} \\  
$V_5$   &  $\bar \alpha_{5/15}$   &   31 &    0.30 & 1.6&                   0.30 &  1.7 & \cite{AAH92}  \\  
$L_{c,kpc}$   &  $L_{320}$        &   28 &    0.63 &$2.3\times 10^{-4}$&    0.31 &  1.9        &  \cite{FTE84} \\  
$m_{opt}$          &  $V_5$       &   29 &    0.33 & 1.3   &                0.31 &  2.0        &  \cite{VTU92} \\  

\sidehead{Correlations likely due to redshift bias}
$L_{opt}$          &  $\beta_{app}$          &   20 &    0.38 & 2.0 &    0.27 & \phn 4.4  & \n \\  
$L_{opt}$          &  $L_{c,pc}$  &   24 &    0.58 & $8.4\times 10^{-3}$ &    0.28 &\phn  6.0  & \n \\  
$L_{opt}$          &  $L_5$      &   30 &    0.54 & $2.4\times 10^{-3}$ &    0.20 & 13 & \n \\  
$L_{c,pc}$    &  $L_{320}$          &   22 &    0.56 & $2.8\times 10^{-2}$ &    0.19 & 22  & \n \\  
$L_{320}$          &  $L_{xray}$          &   23 &    0.48 & $1.4\times 10^{-1}$ &    0.07 & 65  & \n \\  
$L_{opt}$          &  $L_{320}$          &   29 &    0.36 & $5.4\times 10^{-1}$ &    0.02 & 86  & \n \\  
\sidehead{Rejected correlations}
$|\chi_{15} - PA_{pc}|$    &  $EW$       &   20 &    0.40 & 1.5& 0.42 &$3.8\times 10^{-1}$& Due to single outlier \\  
$m_{opt}$   &  $ \bar \alpha_{5/15}$&      29 &  0.32 &  1.6 &   0.32 & 1.6 &  Due to single outlier \\  
$m_C$       &  $\beta_{app}$   &          19 &  0.39  &  1.9 &    0.38 &  1.8 &  Due to single outlier \\  
\enddata 
%\tablenotetext{a}{http: etc...}
\tablecomments{Columns are as follows: (1) and (2) Source property
(see Table~\ref{quantities}); (3) Number of data pairs; (4) Kendall's
tau coefficient; (5) Probability of correlation arising by chance in
per cent; (6) Kendall's tau coefficient, with redshift partialed out;
(7) Probability of correlation arising by chance, with redshift
partialled out; (8) Comments and/or reference for previous detection
of correlation, if applicable.}
 
 \end{deluxetable}
\begin{deluxetable}{lllllll}
%\tabletypesize{\scriptsize}
\tablecolumns{7}
\tablecaption{\label{ks_results}Two-Sample Tests for Various Subclasses of the FS-PR Sample}
\tablewidth{0pt}
\tablehead{ & \colhead{QSO} & \colhead{HPQ} & \colhead{IDV II} & \colhead{IDV I \& II}  & \colhead{IDV II}    & \colhead{2-sided}    \\
 &  \colhead{vs.} &  \colhead{vs.}   & \colhead{vs.} &  \colhead{vs.} &  \colhead{vs.} &  \colhead{vs.}   \\
\colhead{Property} & \colhead{BL} & \colhead{BL}  & \colhead{IDV 0 \& I} & \colhead{IDV 0}  & \colhead{IDV 0}  & \colhead{Others}    \\
\colhead{(1)} & \colhead{(2)} &\colhead{(3)} & \colhead{(4)} &\colhead{(5)} & \colhead{(6)} &\colhead{(7)}  }
\startdata
$z$              &$   0.070  $&$ 1.9      $  &   \n&   \n&   \n&      \n \\  
$EW $       &$   0.0053 $&$ 0.017  $&   \n&$ 0.13  $&$ 0.16  $&$ 0.021$ \\  
$\bar \alpha_{5/15}$   &     \n&      \n&   \n&  $ 0.46  $&$ 0.40  $&   \n     \\  
$V_5$         &     \n&   \n&   $ 0.17  $&   \n&$ 0.077  $&      \n \\  
$V_{15}$          &     \n&   \n&     \n&   \n&   \n&     \n \\  
$L_{320}$     &$   0.16 $&$ 0.50  $&   \n&$ 0.90 $&$ 1.2     $&      \n \\  
$L_5$       &$   0.035  $&$ 0.48  $&   \n&$ 0.25  $&$ 1.7     $&    \n \\  
$L_{opt}$          &$   1.4      $&   \n&   \n&   \n&   \n&    \n \\  
$L_{xray}$          &$   0.31  $&   \n&    \n&   \n&   \n&     \n \\  
$L_{c,pc}$   &     \n&   \n&   \n&    \n&      \n&    \n \\  
$L_{c,kpc}$    &$   0.078  $&$ 0.098  $&   \n& \n&   \n&     \n \\  
$L_{ext}$         & 1.72 &   \n&   \n&      \n&      \n&$ 0.011 $ \\  
$R_{pc}$   &\n &\n&$ 0.19     $& \n & 0.082  &  $ 0.46     $ \\  
$R_{kpc}$      &     \n&      \n&   \n&    \n&   \n&$ 0.098  $ \\  
$m_{opt}$          &$   0.78  $&   \n&$ 0.35  $&$ 0.27  $&$ 0.22  $&      \n \\  
$\Delta PA$    &     \n&   \n&   \n&    \n&   \n&$ 0.54  $ \\  
$T_b$ & \n & \n & \n & 1.98 & \n & \n \\
$\beta_{app}$        &     \n&   \n&     \n&   \n&   \n&    \n \\  
%$\Sigma_{bend}$      &     \n&   \n&     \n&   \n&   \n&     \n \\  
%$N_{bend}$       &     \n&   \n&   \n&     \n&   \n&     \n \\  
$m_5$            &     \n&   \n&   \n&    \n&   \n&     \n \\  
$m_{15}$            &     \n&   \n&   \n&     \n&   \n&      \n \\  
$|\chi_{15} - PA_{pc}|$   &     \n&   \n&    \n&     \n&   \n&   \n \\

\enddata 
\tablecomments{All values represent the probability (in per cent) that the subsamples
were drawn from the same population, based on the source property in
question. Those values not listed exceed $2\%$. Columns are as
follows: (1) Source property (see Table~\ref{quantities}); (2) Quasars
vs. BL Lacs; (3) High-optical polarization quasars vs. BL Lacs; (4)
IDV class II AGNs vs. the combined IDV classes 0 \& I; (5) Combined IDV
class I \& II vs. class 0; (6) IDV class II vs.  class 0; (7) Sources
with two-sided kiloparsec-scale morphology vs. all others.}
 
 \end{deluxetable}

\end{document}